\documentclass[sigconf]{acmart}
\AtBeginDocument{%
  }

\usepackage{listings}

\usepackage{subfigure}

\definecolor{gray50}{gray}{.5}
\definecolor{gray40}{gray}{.6}
\definecolor{gray30}{gray}{.7}
\definecolor{gray20}{gray}{.8}
\definecolor{gray10}{gray}{.9}
\definecolor{gray05}{gray}{.95}

\newlength\Linewidth
\def\findlength{\setlength\Linewidth\linewidth
	\addtolength\Linewidth{-4\fboxrule}
	\addtolength\Linewidth{-3\fboxsep}
}

\newenvironment{rqbox}{\par\begingroup
	\setlength{\fboxsep}{5pt}\findlength
	\setbox0=\vbox\bgroup\noindent
	\hsize=0.95\linewidth
	\begin{minipage}{0.95\linewidth}\normalsize}
	{\end{minipage}\egroup
	\textcolor{gray20}{\fboxsep1.5pt\fbox
		{\fboxsep5pt\colorbox{gray05}{\normalcolor\box0}}}
	\endgroup\par\noindent
	\normalcolor\ignorespacesafterend}

\setcopyright{none}
\renewcommand\footnotetextcopyrightpermission[1]{}
\renewcommand\acmConference[4]{}

\begin{document}

\pagestyle{plain}

\title{Automated Test Suite Enhancement Using Large Language
Models with Few-shot Prompting}

\author{Alex Chudic}
\email{alex@freetobook.co.uk}
\orcid{0009-0004-0067-1477}
\affiliation{%
  \institution{US Booking Services Ltd. (freetobook)}
  \country{United Kingdom}
}

\author{G\"{u}l \c Cal{\i}kl{\i}}
\email{handangul.calikli@glasgow.ac.uk}
\orcid{0000-0003-4578-1747}
\affiliation{%
  \institution{University of Glasgow}
  \country{United Kingdom}
}

\begin{abstract}
Unit testing is essential for verifying the functional correctness of code modules (e.g., classes, methods), but manually writing unit tests is often labor-intensive and time-consuming. Unit tests generated by tools that employ traditional approaches, such as search-based software testing (SBST), lack readability, naturalness, and practical usability. LLMs have recently provided promising results and become integral to developers’ daily practices. Consequently, software repositories now include a mix of human-written tests, LLM-generated tests, and those from  tools employing traditional approaches such as SBST. While LLMs’ zero-shot capabilities have been widely studied, their few-shot learning potential for unit test generation remains underexplored. Few-shot prompting enables LLMs to learn from examples in the prompt and automatically retrieving such examples could enhance test suites. This paper empirically investigates how few-shot prompting with different test artifact sources, comprising human, SBST, or LLM, affects the quality of LLM-generated unit tests as \textit{program comprehension artifacts} and their contribution to improving existing test suites by evaluating not only correctness and coverage but also readability, cognitive complexity, and maintainability in hybrid human–AI codebases. We conducted experiments on HumanEval and ClassEval datasets using GPT-4.o, which is integrated into GitHub Copilot and widely used among developers. We also assessed retrieval-based methods for selecting relevant examples. Our results show that LLMs can generate high-quality tests via few-shot prompting, with human-written examples producing the best coverage and correctness. Additionally, selecting examples based on the combined similarity of problem description and code consistently yields the most effective few-shot prompts.\\
\textbf{Data and Materials}: \url{https://doi.org/10.5281/zenodo.15561007}
\end{abstract}


\maketitle


\section{Introduction}
\label{sec:intro}

Software testing is a vital stage of the software development life cycle, ensuring the software satisfies high-quality standards and works reliably. Its main objective is to detect bugs and identify regression early in development. Doing so minimises the cost associated with maintenance and reduces the risk of post-deployment failures. Unit testing is a cornerstone of software quality assurance, focusing on verifying the correctness of individual software components in isolation. 

Beyond verification, unit tests serve as \textit{executable documentation} that developers rely on to understand program behavior, design intent, and edge cases, particularly during maintenance and evolution. As a result, the quality and structure of test code directly affect program comprehension. Ensuring coverage across all software components often requires a large number of test cases. Developers have traditionally written these test cases by hand, but this approach is becoming less popular because of its laborious and time-consuming nature. Given the inherent complexity of modern software systems, manually creating and maintaining high-quality unit tests is inefficient. As a result, companies are pushing towards adopting automated methods to generate and maintain a test suite efficiently.

Driven by the need for more efficient and scalable testing solutions, automated unit testing has become one of the main topics among researchers in the field of software engineering. The traditional approaches, such as random-based \cite{pacheco_randoop_2007, pacheco_feedback-directed_2007}, constraint-based \cite{xiao_characteristic_2013}, and search-based testing \cite{fraser_evosuite_2011, harman_theoretical_2010, harman_search-based_2012}, have demonstrated their ability to generate tests with reasonable test coverage. However, the created test cases lack readability, and professional developers find them less comprehensible compared to those created manually \cite{alagarsamy_a3test_2023}. These shortcomings pose a major obstacle for adoption and usability across different environments, making it particularly difficult for beginner developers.

Over recent years, test generation tools have evolved significantly with the substantial advancements in machine learning and large language models (LLMs). Following the discovery of Vaswani et al. \cite{vaswani_attention_2023}, numerous deep-learning tools were built to address the limitations of the traditional methods. AthenaTest \cite{tufano_unit_2021} was one of the first models that employed the domain adaptation principle of transformers, adapting the knowledge of the model pre-trained on a large corpus of code to the purpose of unit test generation. Another deep-learning tool, A3Test \cite{alagarsamy_a3test_2023}, improves this approach further by providing assertion knowledge and verifying naming consistency, which leads to increased correctness and code coverage. Despite the improved results, the deep-learning tools generate a relatively low percentage of valid test cases (16.21\% and 40.05\% respectively), with poor line coverage of the focal methods.

Most recently, with the emergence of highly versatile LLMs, models such as ChatGPT \cite{openai_gpt-4_2023} and Codex \cite{chen_evaluating_2021} revolutionised a wide range of natural language processing (NLP) and code generation tasks. They are trained on vast datasets of natural language and code repositories, which makes them proficient in NLP and code-related tasks. Numerous studies introducing LLM-based test generation tools have been published \cite{fan_large_2023, wang_software_2023}, highlighting superior results in terms of readability of generated code. However, hallucination (generating reasonable but incorrect code) and non-determinism (producing varying outputs for identical prompts) remain the key challenges that need to be addressed. As a result of hallucination, a high proportion of the LLM-generated test cases fail due to compilation or execution errors, in most cases caused by minor issues, e.g., syntactic issues or missing imports \cite{siddiq_using_2024, yuan_no_2024}. Tailored post-processing techniques, such as test case repair or interactive prompt improvement, are shown to mitigate this issue by fixing the affected components \cite{xie_chatunitest_2023, siddiq_using_2024, schafer_adaptive_2023, yuan_no_2024}. When it comes to non-determinism, the temperature hyperparameter controls the randomness to a limited extent. Even at the lowest setting, it does not guarantee fully deterministic behaviour, which complicates the reproducibility of empirical studies \cite{ouyang_llm_2023}. However, when managed strategically via comprehensive validation and improvement assessment, non-determinism can foster creativity, which can be beneficial for targeting edge cases and uncommon execution paths \cite{doderlein_piloting_2023, alshahwan_automated_2024}.

The effectiveness of the LLMs in unit test generation heavily depends on how the task is formulated in the input prompt, which is formally called prompt engineering. The context provided in the prompt can significantly affect the coverage and code quality of the generated test-cases \cite{yang_evaluation_2024, siddiq_using_2024}. Advanced prompt engineering approaches, such as few-shot prompting and chain-of-thought (CoT), have been shown to enhance the correctness of the generated test cases by providing structured examples or explicit reasoning steps \cite{brown_language_2020, ouedraogo_large-scale_2024}. In this study, we will focus on few-shot learning, a technique where example test cases are included in the prompt to demonstrate expected outputs and guide the model towards generating higher-quality test cases.

A key consideration in few-shot prompting is the origin of the example test cases, whether they are human-written, generated using search-based software testing (SBST), or LLMs. Our work examines how various example sources impact the code quality and comprehensibility of LLM-generated test cases and their effectiveness in enhancing the existing test suite. This is a practical problem in software development, as projects often start with a partial test coverage comprised of test cases that can be created by human developers or generated by different tools. The goal is to generate high-quality, coverage-improving, and human-comprehensible unit tests that complement rather than duplicate the existing test cases, facilitating software maintainability and evolution.

Furthermore, using relevant code demonstrations in few-shot learning is shown to be effective in different code-related tasks \cite{noor_retrieval_2023}. Therefore, our study compares the use of retrieval-based methods for selecting few-shot examples against random-based methods. We evaluate the impact of selection methods on code quality and the practical utility of LLM-generated test cases. By systematically analysing these factors, we aim to establish good practices for prompt engineering in unit test generation, balancing creativity and precision to maximise the coverage and maintainability of the enhanced test suite.

Our review reveals a critical research gap in understanding how demonstration quality and different selection techniques affect the quality of the generated code. To address this, our study will evaluate few-shot learning using different test case sources (human-written, SBST-generated, and LLM-generated) with various selection techniques, as detailed in Section~\ref{sec:methodology}. We aim to bridge the gap and build upon prior work~\cite{ouedraogo_large-scale_2024, noor_retrieval_2023} to understand the practical implications of using existing test suites as example sources and how example selection affects the quality of test generation. By measuring the coverage improvement of the newly generated test cases, we aim to provide actionable insights into the real-world utility of LLM-generated tests for enhancing test suites.

In the next section, we introduce the prompting techniques and review the related literature on LLM test generation. In Section \ref{sec:methodology}, we motivate and outline our research efforts anddescribe the design of our experimental setup. In Section \ref{sec:results}, we present the analysis of the evaluation results. In Section \ref{sec:conclusion}, we summarise the findings, describe the limitations of our approach, and propose avenues for future work.

\section{Background and Related Work}
\label{sec:background}

In this section, we introduce prompt engineering techniques employed in unit-test generation and present existing studies that use these prompting techniques to generate unit tests with or without augmenting traditional automated test generation tools in the form of prototypes or evaluation pipelines integrated into software practitioners' workflows. 

\subsection{Prompt Engineering Techniques for Unit Test Generation}
\label{sec:prompt_engineering}

Prompt engineering refers to constructing the optimal prompt that yields the best outputs for a given task.  
The recent LLM-based test generation studies rely on three main prompting practices: zero-shot prompting, few-shot prompting, and chain-of-thought (CoT) reasoning.

\noindent \textbf{Zero-shot prompting} is the most widely-used approach in LLM-based unit test generation~\cite{wang_software_2023} and refers to the most straightforward prompting technique when the message merely conveys what we want the model to do. Sch\"{a}fer et al. \cite{schafer_adaptive_2023} introduced TESTPILOT, an automated JavaScript test generator using Codex \cite{chen_evaluating_2021}, which employs iterative prompt refinement by re-prompting failed tests with error messages for repair. Evaluated on 25 npm packages, TESTPILOT achieved 68.2\% statement coverage and 47.1\% correctness. ChatTester, by Yuan et al. \cite{yuan_no_2024} also applied iterative prompt refinement for test repair, improving test correctness from 22.3\% to 41\%. Recent work explored hybrid repair strategies for failed test cases. ChatUnitTest by Xie et al. \cite{xie_chatunitest_2023} combined rule-based and ChatGPT-based repairs, achieving 29.98\% correctness and 89.36\% line coverage on Defects4J and other Java projects. Siddiq et al. \cite{siddiq_using_2024} evaluated Codex, GPT-3.5, and StarCoder with heuristic repair, noting wide performance variation: up to 87.4\% coverage on HumanEval but under 2\% on EvoSuite SF110. These differences underscore the need for diverse benchmarks to assess generalizability of proposed LLM-based unit test generation approaches.

Overall, LLMs show strong potential for automated test generation, outperforming tools like AthenaTest and A3Test in coverage and correctness. Yet, many generated tests still contain execution-blocking errors, highlighting the importance of integrating test repair methods—whether rule-based or iterative—to improve reliability.

\noindent \textbf{Few-shot prompting} allows LLMs to learn at runtime from examples input in the prompts, offering a clear structure for the output to follow \cite{brown_language_2020}. 
The quality of the provided examples significantly impacts the effectiveness of few-shot prompting. Providing high-quality test examples ensures the LLM generates well-structured outputs \cite{ouedraogo_large-scale_2024}. However, manually crafting optimal examples can be a challenging task. In recent work, this subject has been addressed by employing retrieval-based example selection for program repair and test assertion generation tasks \cite{noor_retrieval_2023}, demonstrating that retrieved contextually relevant examples significantly improve output accuracy compared to randomly selected ones. In the domain of test generation, few-shot learning has been shown to generate test cases with superior readability and maintainability in comparison with zero-shot and CoT methods, hinting at its potential for generating high-quality test cases \cite{ouedraogo_large-scale_2024}. Retrieval-based example selection for few-shot prompts can facilitate high-quality LLM-based unit test generation.

The study by Ou\'{e}draogo et al. \cite{ouedraogo_large-scale_2024} evaluated five prompt engineering techniques, including few-shot prompting for unit test case generation without any test error-resolution framework. The results show minimal test correctness and underwhelming line coverage across all prompting methods, with negligible differences. However, the authors found that the prompts following a few-shot learning design led to the fewest syntax violations, indicating enhanced readability and maintainability. In their study, the example used for the few-shot learning is selected manually, and it is consistently used across all experiments. The authors suggest that good input examples should translate into well-structured LLM outputs. These findings underscore the potential of few-shot learning for generating high-quality test cases.

Research by Harman et al. \cite{alshahwan_automated_2024} explored industrial applications of LLMs for test generation. The authors used TestGen-LLM, a custom LLM from Meta, to expand the existing Kotlin test suite for Meta platforms. They employed few-shot prompts containing a single test case example to generate complementary test and corner cases. By utilising a pipeline (i.e., Assured LLM-based Software Engineering pipeline) to reliably filter the correct and coverage-improving test cases, the study reported 25\% of generated test cases passing this filter, while maintaining a high acceptance rate (73\%) for production deployment.

\noindent \textbf{Chain-of-thought} (CoT) encourages the model to apply complex reasoning by following a step-by-step thought process~\cite{wei_chain--thought_2023}. CoT extends the standard employed prompting strategy (i.e. zero-shot or few-shot) by including an explicit reasoning trigger, e.g. \textit{"Let's think step by step"}, before the task definition. Research suggests that CoT is particularly useful for solving multi-step arithmetic and logical reasoning tasks, which require structured reasoning \cite{kojima_large_2023}. For LLM test generation, CoT demonstrated mixed results. Yang et al. \cite{yang_evaluation_2024} report that it improved the effectiveness of the DeepSeek-Coder models in unit test generation, but it yielded negligible improvement for CodeLlama. Notably, Oue\'{e}draogo et al. \cite{ouedraogo_large-scale_2024} demonstrate that CoT yielded the best coverage (55.64\%) using GPT-3.5 without any test error-correction methods, showing the potential of CoT in LLM-based test generation. 

\subsection{Augmenting Traditional Tools with LLMs}
Recent work explores combining LLMs with traditional automated testing to improve coverage by targeting untested paths and edge cases. Liu et al.~\cite{liu_is_2023} introduce EvalPlus, which augments existing test suites using zero-shot LLM generation and mutation-based testing, achieving a 28.9\% increase in error detection on HumanEval. However, reliance on HumanEval’s simple problems raises concerns about generalisability to more complex systems.

Lemieux et al.~\cite{lemieux_codamosa_2023} propose CODAMOSA, a hybrid framework integrating Pynguin SBST with LLMs to overcome coverage plateaus by generating tests for uncovered functions. CODAMOSA achieves 10–15\% higher coverage than SBST and LLM-only baselines, but its effectiveness depends on the quality of LLM-generated tests, which varies with prompt design and program complexity.

\subsection{Assured LLM-based Software Engineering}
Prior work shows that inherent LLM limitations (e.g., hallucination and non-determinism) hinder test generation and proposes mitigation strategies such as iterative prompt refinement and test repair. Harman et al.~\cite{alshahwan_assured_2024} formalize these efforts through Assured LLM-based Software Engineering (Assured-LLMSE), which guarantees preservation of existing functionality and measurable improvement. They propose a generate–verify–validate pipeline that checks test executability and evaluates utility using predefined metrics, mitigating common LLM failures. Our evaluation pipeline is inspired by Assured-LLMSE and aligns with prior LLM-based test generation tools~\cite{alshahwan_automated_2024, xie_chatunitest_2023}.
\section{Methodology}
\label{sec:methodology}
This section formulates the research questions and provides details about the experimental setup. During our experiments, we used GPT-4.o due to its integration to GitHub CoPilot and demonstrated performance of GPT models in various research studies on unit test generation~\cite{schafer_adaptive_2023, ouedraogo_large-scale_2024, xie_chatunitest_2023}. 

\subsection{Research Questions}

This paper focuses on the enhancement of existing test suites using few-shot learning for unit-test generation for the following reasons: \textbf{(i)} few-shot learning has been far less explored than unit-test generation using zero-shot learning; \textbf{(ii)} it provides more readable and maintainable unit tests compared to zero-shot learning and CoT. According to the previous work by Ou\'{e}draogo et al. \cite{ouedraogo_large-scale_2024}, the correctness and coverage of unit tests generated using few-shot learning were not the highest. However, the authors used manually created examples in the few-shot prompts, which were the same for the unit test suite generation of all Java classes. Recent work on leveraging LLMs for program repair and test assertion generation showed that automated (retrieval-based) selection of contextually relevant examples significantly improves the outcome \cite{noor_retrieval_2023}. Therefore, due to such promising results obtained in related software development tasks, we focus on the automated retrieval of examples from existing unit tests to overcome the challenge of manually crafting optimal examples to include in few-shot prompts and generate improved quality unit tests.  

Software development repositories can be hybrid regarding the source that implemented/generated the unit tests: (human) developers could have implemented some unit tests, whereas the rest could have been generated by tools that employ traditional approaches (e.g., search-based testing). With the emergence of LLMs, software repositories now also contain unit tests generated by leveraging LLMs. Unit tests manually crafted by developers and generated through an automated approach have different characteristics. For instance, tests automatically generated through search-based methods can improve text coverage, however they can also lack readability~\cite{roy_deepTC_2021}, naturalness, and practical usability~\cite{panichella_testsmells_2020}. Using test examples from various sources (developers, unit test generation tools, LLMs) in few-shot prompts can impact the quality of the LLM-generated unit tests. Therefore, to investigate such a possibility, we formulate our first research question as follows: 

\begin{center}
    \begin{rqbox}	    
       \textbf{RQ$_1$.} \emph{How does using test examples implemented/generated by different sources (i.e., developers, SBST and LLMs) in few-shot prompts affect the quality of LLM-generated tests?}
    \end{rqbox}
\end{center} 
 
Moreover, including relevant example unit tests in the few-shot prompts can improve the quality of LLM-generated unit tests compared to using randomly selected examples in the prompts. Therefore, our second research question investigates the impact of automated retrieval-based selection of contextually relevant examples on the generated unit tests' quality.

\begin{center}
    \begin{rqbox}	    
       \textbf{RQ$_2$.} \emph{How does the use of retrieval-based techniques for selecting few-shot unit test examples affect the quality of LLM-generated tests?}
    \end{rqbox}
\end{center}

Our study focuses on enhancing existing test suites using few-shot learning, aiming to provide insights into LLM-based unit test generation for real-life software repositories with test suites that need enhancements. Hence, through RQ3, we want to investigate the impact of using test examples from different sources on the improvement of existing test suites’ quality (i.e., coverage, correctness, code quality).

\begin{center}
    \begin{rqbox}	    
       \textbf{RQ$_3$.} \emph{How does using test examples implemented/generated by different sources (i.e., developers, SBST and LLMs) in few-shot prompts affect enhancement of the existing test suites' quality?}
    \end{rqbox}
\end{center}

Furthermore, we want to investigate the impact of automated retrieval-based selection of contextually relevant examples on the improvement of existing test suites’ coverage, which corresponds to our last research question RQ4 that is formulated as follows: 

\begin{center}
    \begin{rqbox}	    
       \textbf{RQ$_4$.} \emph{How does using retrieval-based techniques to select few-shot unit test examples affect enhancement of the existing test suites' quality?}
    \end{rqbox}
\end{center} 

\subsection{Datasets}
\label{subsec:benchmarks}
As highlighted in prior research \cite{siddiq_using_2024}, the effectiveness of LLM test generation can vary significantly on different benchmarks. Therefore, our experiments use HumanEval~\cite{chen_evaluating_2021} and ClassEval~\cite{du_classeval_2023} datasets, offering various problem complexities. While HumanEval is a simple code benchmark consisting of isolated coding exercises, ClassEval tasks contain cross-file references that require tests with class-level references to achieve high coverage. This combination of benchmarks facilitates a comprehensive examination of LLMs' ability to generate unit tests through few-shot prompting. 
Both benchmarks include human-developed unit tests, which are crucial to answering RQ1. Answering RQ1 requires a dataset comprising human-written tests, SBST-generated and LLM-generated tests. The test cases across HumanEval and ClassEval datasets vary in format. Therefore, to form human-written tests, we converted test cases in these two benchmarks into \texttt{pytest}. We generated 211 SBST-based unit tests for the HumanEval dataset and 229 SBST-based unit tests for the ClassEval dataset, using Pynguin~\cite{noauthor_pynguinpython_nodate}. Finally, we created the dataset comprising LLM-generated tests by employing GPT-4o, asking it to generate unit tests for the model solutions extracted from the benchmarks using a simple prompt inspired by prior studies \cite{xie_chatunitest_2023, ouedraogo_large-scale_2024}. For the HumanEval dataset, the total number of LLM-generated unit tests is 1754, while we generated 1259 LLM-based unit tests for the ClassEval dataset.

\subsection{Measuring the Quality of LLM-generated Unit-Tests}
\label{subsec:unit-test_quality_metrics}
To answer Research Questions RQ1 and RQ2, we need to assess the quality of the generated unit tests. However, comprehensive test quality evaluation is a complex task that a single metric cannot fully capture. Therefore, we employ a diverse set of metrics to evaluate various aspects of test quality referring to established approaches in test generation research \cite{wang_software_2023, ouedraogo_large-scale_2024, siddiq_using_2024}. We organized these metrics into three main categories as follows: 

\noindent \textbf{Functional correctness} reflects the practical usability of the generated tests, as the test cases which fail or contain faulty syntax are not suitable for use. Our assessment considers three metrics widely used in the literature \cite{liu_is_2023, yuan_no_2024, nie_learning_2023}: 
\begin{itemize} 
    \item \textit{Syntactic Correctness} validates test case syntax using an Abstract Syntax Tree (AST) parser.
    \item \textit{Compilation Correctness} determines whether the generated tests compile successfully.
    \item \textit{Execution Correctness} checks whether generated tests pass without errors when executed.
\end{itemize}

\noindent \textbf{Code coverage} is a fundamental metric that measures the proportion of the code executed by the test suite. The higher coverage generally indicates more thorough testing, but does not guarantee fault detection. To evaluate the coverage, we will use the two widely adopted metrics \cite{xie_chatunitest_2023, siddiq_using_2024, zhang_assessing_2024}, which will be computed using a combination of established Python testing modules - \texttt{pytest} and \texttt{coverage}:
\begin{itemize}
    \item \textit{Line coverage} measures the percentage of executable statements in the code base that have been traversed at least once during the test suite execution.
    \item \textit{Branch coverage} calculates the proportion of decision branches that have been executed during testing. A branch is considered covered if both true and false components have been executed at least once during testing.
\end{itemize}

Code coverage is widely used as an indicator of test suite effectiveness \cite{zhang_assessing_2024, ahmed_can_2016}, though its correlation with actual defect detection depends on factors like test quantity \cite{inozemtseva_coverage_2014}. Since effectiveness is multifaceted—affected by assessment quality, redundancy, and more—and lacks a standard metric, we adopt coverage as a practical proxy in this study while acknowledging its limitations.

\noindent \textbf{Test code quality} represents another critical dimension of the test suite quality, with direct implications for program comprehension, maintenance costs and its ability to find defects. Research shows that poor test quality, often manifested through test smells and excessive complexity, can significantly degrade the long-term value of the test suite while increasing technical debt \cite{tufano_when_2015, bavota_empirical_2012}. This research question employs four established metrics to assess test quality:

\begin{itemize}
    \item \textit{Cyclomatic complexity} assesses the complexity of the test case from the computational point of view by calculating the number of independent execution paths through a program. High complexity suggests low performance, and can identify overly complex or redundant test cases.
    \item \textit{Cognitive complexity} measures how challenging it is for the developer to understand the test case. Unlike cyclomatic complexity, this metric focuses on factors like structure and naming. Lower cognitive complexity leads to better readability and maintainability.
    \item \textit{Test smells} represent the common anti-patterns in unit tests, such as excessive setup, hard-coded values, or duplicate code. These "smells" indicate poor programming practices, which often reduce maintainability and introduce extensive complexity. Recent studies in unit test generation \cite{siddiq_using_2024, ouedraogo_large-scale_2024} have validated test smell as a robust quality indicator, demonstrating its predictive value for identifying maintainability issues.  
    \item \textit{Maintainability} measures the effort required to maintain a test suite over time, considering factors such as code readability, complexity, and duplication. 
\end{itemize}
 To compute  cyclomatic and cognitive complexities and detect code smells, we used SonarScanner CLI~\cite{sonarscanner-cli}, which performed static analysis  (i.e., parsed source and test code and applied language-specific analysis rules). The resulting code quality metrics were uploaded to SonarCloud, which was used to store, aggregate, and report analysis results (e.g., number of test smells). Test files were explicitly identified via scanner configuration, allowing all metrics to be scoped to unit test code only. We operationalized test smells as code smell issues reported in test files and extracted test-specific technical debt using SQALE rating~\cite{sqale-rating} of SonarQube Cloud~\cite{sonarqube-cloud}. Cognitive and cyclomatic complexity were used as proxies for the effort required to comprehend unit tests, following established practice in program comprehension research.

The selected metrics offer a well-rounded evaluation of LLM-generated tests—beyond correctness and coverage—by also assessing readability, maintainability, and robustness, enabling meaningful comparisons across few-shot prompting methods. This comprehensive approach aligns with prior work emphasizing diverse metrics for realistic test quality assessment \cite{yuan_no_2024, chen_evaluating_2021}.

\subsection{Prompt Design}
\label{sec:prompt_design}

 We employ several prompt engineering techniques introduced in Section~\ref{sec:prompt_engineering} alongside few-shot prompting to leverage the LLM capabilities fully. The prompt comprises two components: system prompt and user prompt.

\textit{System prompt} defines the behaviour we expect from the model (Figure \ref{fig:system_prompt}). We utilize the Persona Pattern \cite{white_prompt_2023} to clearly describe the role the model should assume, which is widely used in similar problems \cite{yuan_no_2024, ouedraogo_large-scale_2024}. Additionally, specifying the key details and metrics helps ensure the output of the LLM aligns with the intended objectives. Furthermore, by referencing $EXAMPLES$ within the system message, we utilize the Few-shot Code Example Generation Pattern \cite{white_chatgpt_2023}, 
which helps the model recognize these examples in the user prompt by the same reference.

\begin{figure}[t]
    \centering
    \setlength{\fboxsep}{10pt}
    \fbox{\parbox{0.45 \textwidth}{
        \centering
        You are an expert in Python test generation using pytest. Your goal is to generate new high-quality unit tests for a given Python class. You will be provided with the class definition and your output should be a list of new unit tests. The prompt will include EXAMPLES of similar test cases to help you generate well-structured test cases. Make sure to keep the tests maintainable and easy to understand, while aiming for high code coverage. The output should only include the test classes.
    }}
    \caption{System prompt}
    \label{fig:system_prompt}
\end{figure}

\begin{figure}[t]
    \centering
    \begin{lstlisting}
# CLASS UNDER TEST: {class_name}
{class_content}

# EXAMPLES:
{selected_test_cases_content}
    \end{lstlisting}
    \caption{User prompt}
    \label{fig:user_prompt}
\end{figure}

\textit{User prompt} includes most of the relevant information about the class under test. It consists of the $class\_name$, the whole code definition of the class, and the few-shot learning examples. As shown by Siddiq et al. \cite{siddiq_using_2024}, using the full class content leads to the best resulting test correctness. However, given the information-dense nature of the provided context, we decided to use a straightforward structure to maintain clarity (Figure \ref{fig:user_prompt}), referring to prompt structures commonly used in related literature \cite{xie_chatunitest_2023, yuan_no_2024, ouedraogo_large-scale_2024}, where similar approaches have been effective.

\noindent \textbf{The Number of Examples.} To determine the optimal number of test examples ($n_{examples}$) to include in the prompts, we conducted preliminary experiments with three prompt configurations ($n_{examples} = 1, 3,$ and $5$). According to the results of the preliminary experiments, (branch and line) coverage for both human- and LLM-based examples improves significantly when $n_{examples} = 5$,  although there is a slight decrease in coverage for SBST-based examples compared to the case when $n_{examples} = 1$. We observed no difference in the percentages of syntactically correct and compilable tests across the 1, 3, and 5 example settings. However, we observed a significant increase in the test pass rate for human-written test examples with $n_{examples} = 5$, while a slight decrease for the SBST and LLM-based test examples. Finally, we did not observe any clear pattern for the code quality metrics. Our findings are consistent with with prior findings \cite{noor_retrieval_2023}. Therefore, we use five-example few-shot prompting in this study. The results of the preliminary experiment are presented in the artifacts package\footnote[1]{Please refer to sub-folder \texttt{/data/figures} in the artifacts package that is available on Zenodo: \url{https://doi.org/10.5281/zenodo.15561007}}.

\begin{figure*}
    \centering
    \includegraphics[width=2\columnwidth]{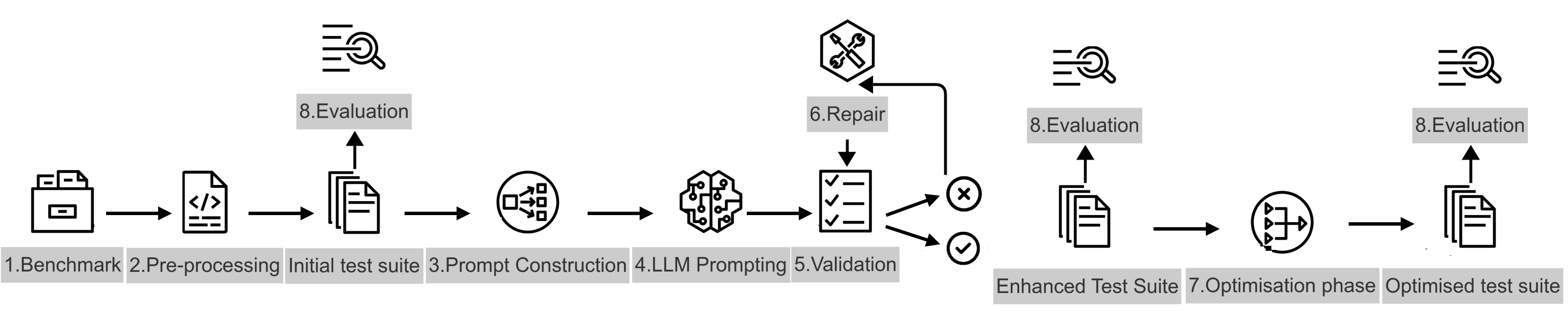}
    \caption{High-level overview of the evaluation pipeline. Stage ``8. Evaluation" in the pipeline comprises evaluation of test suites' \textit{functional correctness}, \textit{coverage} and \textit{code quality}. We conduct evaluation on the (i) initial test suite (after stage ``2. Pre-processing"); \textbf{(ii)} enhanced test suite (after stages ``5. Validation" and ``6. Repair") and \textbf{(iii)} optimised test suite (after stage ``7. Optimisation phase").}
    \label{fig:pipeline-chart}
\end{figure*}

\subsection{Techniques to Retrieve Test Examples}
\label{subsec:test-selection-methods}

Retrieval-based techniques (e.g. cosine similarity, TF-IDF) have proven effective in diverse code-related tasks by identifying contextually relevant items \cite{noor_retrieval_2023}. By quantifying the similarity between classes, they can help identify the most relevant classes from which to draw test examples. In RQ2, we systematically evaluate how retrieval-based selection methods, based on the similarity of different class features, compare to random baseline methods in generating high-quality tests. The unit test example selection methods we will consider are the following:

\noindent \textbf{Random Choice from Test Suite}. This approach randomly selects unit tests from the entire test suite. This technique will serve as a great baseline to compare against, as it disregards all available context.

\noindent \textbf{Random Choice from Class under Test}. The second random approach also chooses randomly, although this time from the class under test (CUT). Therefore, any of the selected unit tests will be highly relevant to the CUT, providing valuable context for the CUT. However, inputting the related unit test may bias the LLM towards a specific type of test, which can lead to repetitiveness, resulting in generated tests that do not improve the overall coverage of the test suite.

\noindent \textbf{Similarity-based Selection Techniques}. These selection methods convert problem description (in text format) and code into vectors using TF-IDF and calculate the textual similarities using cosine similarity. We select the most relevant examples through textual similarities based on similarities between classes or problem definitions.
\begin{enumerate}
    \item \textit{Problem definition similarity}. Finds the test cases whose classes have the most similar problem definition, i.e. the natural language definition of what the class should do.
    \item \textit{Source code similarity}. Chooses test cases from the classes with the most similar source code.
    \item \textit{Source code similarity with problem definition comment}. The benchmark solution source code includes a comment at the top of the file, describing how the function or class should work in natural language. This method includes this comment in determining the similarity between class source code.
    \item \textit{Problem and source code similarity}. Aggregates the scores from \textit{Problem definition similarity} and \textit{Source code similarity} to select the most similar test cases.
\end{enumerate}

\subsection{Test Repair}
\label{subsec:repair_rules}

Our preliminary analysis, which focused on the functional correctness of the generated tests, revealed significant correctness problems. This finding aligns with the results from related literature \cite{schafer_adaptive_2023, yuan_no_2024, xie_chatunitest_2023, siddiq_using_2024, alshahwan_automated_2024}. To prevent discarding a large number of generated tests, we implemented rule-based repairs for common failure problems, following methods employed in prior research \cite{xie_chatunitest_2023, siddiq_using_2024}. 
Our rule-based repair mechanism targets the most common issues observed in the validation stage, ensuring that the generated test cases are reliable and suitable for subsequent evaluation. The following rules were defined to automatically address the identified issues:

\noindent \textit{Rule 1: Handling incomplete outputs}. Incomplete test cases are removed due to the output token limit.

\noindent \textit{Rule 2: Adding missing pytest import}. Occasionally, the generated test cases fail due to the missing pytest import. In such cases, we simply add \texttt{import pytest} at the top of the file. 

\noindent \textit{Rule 3: Adding missing class under test (CUT) import statement}. A large number of test cases fail because the CUT import statement is missing or incorrectly defined. When this occurs, the desired import is added to the top of the file.

\noindent \textit{Rule 4: Removing faulty module import statement}. Sometimes, the corrupted import statement causes compilation errors. We identify these by parsing the error message, and fix them by removing the faulty import statements.

\noindent \textit{Rule 5: Removing self argument from standalone test functions}. Sporadically, standalone functions are incorrectly generated with the \texttt{self} argument. When this is the case, we remove the \texttt{self} parameter to prevent errors.

\noindent \textit{Rule 6: Removing the function definition from the class under test from the test class}. LLM-generated test classes sometimes include the function under test within the test class itself, which can mislead the evaluation. We resolve this by removing the function definition.

\noindent \textit{Rule 7: Add missing function names}. When the LLM-generated test cases consist of plain assert statements, we wrap them in a function. This is required for accurate pytest evaluation. 

\noindent\textit{Rule 8: Removing test cases with \texttt{SyntaxError}}. Occasionally, the LLM-generated test cases contain syntax errors (e.g. missing brackets), which prevent pytest execution. To avoid impacting evaluation, test cases that cause \texttt{SyntaxError} are automatically removed.

\noindent\textit{Rule 9: Removing test cases with no executable code}. LLM-generated tests that contain only comments (without any executable code) cause \texttt{IndentationError}, halting \texttt{pytest} and falsely reporting coverage as zero. To avoid incorrect coverage calculation, we remove these test cases.

\subsection{Evaluation Pipeline}
In this section, we present the design of our evaluation setup, which is used to assess the effectiveness of LLM test generation with few-shot learning. Its goal is to systematically gather code quality data to rigorously answer the RQs defined in Section \ref{sec:methodology}. A high-level summary of the pipeline is illustrated in Figure \ref{fig:pipeline-chart}, with detailed description below:

\noindent \textbf{(1) Benchmark}: First, we select suitable code generation benchmarks that meet our requirements. As explained in Section \ref{subsec:benchmarks}, we use HumanEval and ClassEval to ensure comprehensive analysis.

\noindent \textbf{(2) Pre-processing}: In this step, we convert the benchmark dataset into a suitable format and generate the initial SBST-based and LLM-generated datasets (detailed in Section \ref{subsec:benchmarks}).

\noindent\textbf{(3) Prompt Construction}: Next, one of the example unit test selection methods (described in Section \ref{subsec:test-selection-methods}) is employed to retrieve the set of example unit tests. These examples are then assembled into a prompt following the structure introduced in Section \ref{sec:prompt_design}.

\noindent \textbf{(4) LLM prompting}: In the prompting stage, the constructed prompts are organised into batches and input into the GPT-4o model. Once the batch is finished, the generated tests are then saved into Python files to enable further processing.

\noindent \textbf{(5) Validation}: The validation process ensures the functional correctness of the LLM-generated test cases. This is done in the following stages: (1) \textit{Syntactic correctness} is checked by employing an AST parser on the test source code to ensure it adheres to the syntactic rules. (2) \textit{Compilation correctness} is determined by compiling the tests. This step ensures all dependencies are properly referenced and the import statements are valid. (3) \textit{Execution correctness} is evaluated by observing whether the tests pass without triggering any runtime errors. We attempt to repair the test cases that fail any of the stages.

To handle tests that enter infinite loops, we employed a 2-minute timeout mechanism per test case, inspired by Siddiq et al. \cite{siddiq_using_2024}. When a test surpasses this execution limit, it is terminated and marked as non-functional.

\noindent \textbf{(6) Rule-based repair}: Research literature suggests a low correctness rate of the LLM-generated tests \cite{siddiq_using_2024, yuan_no_2024}. Therefore, we implement rule-based repair that attempts to fix problematic tests. This mechanism is detailed in Section \ref{subsec:repair_rules}.

\noindent\textbf{(7) Test suite optimisation}: The optimisation phase addresses RQ3 and RQ4 by evaluating the LLM-generated test cases and assessing their practical utility. Each test is assessed by measuring its impact on the line and branch coverage of the initial test suite. The coverage-improving tests are retained in the optimised suite, while the redundant tests are discarded to minimise redundancy.

\noindent\textbf{(8) Test suite evaluations}: To assess the impact of few-shot learning examples, we evaluate functional correctness, coverage, and code quality at three stages in the pipeline. First, we establish the baseline measurements using the initial test suite, which comprises the human written test suites that already exist in HumanEval and ClassEval datasets converted to pytest format (see Section~\ref{subsec:benchmarks}). The second evaluation is performed solely using the new LLM-generated test cases using few-shot learning, following validation and rule-based repair, thereby addressing RQ1 and RQ2 by assessing their standalone quality. Finally, we evaluate the optimised test suite (i.e. initial tests with the coverage-improving LLM-generated tests with few-shot prompting) to determine the overall enhancement from our approach.

\section{Results and Analysis}
\label{sec:results}

\begin{table}[t]
    \centering
    \resizebox{\columnwidth}{!}{
    \begin{tabular}{lrrrrrr}
        \toprule
        & \multicolumn{3}{c}{\textbf{HumanEval}} & \multicolumn{3}{c}{\textbf{ClassEval}} \\
        \cmidrule(lr){2-4} \cmidrule(lr){5-7}
        \textbf{Metric} & \textbf{Human} & \textbf{Pynguin} & \textbf{ChatGPT} & \textbf{Human} & \textbf{Pynguin} & \textbf{ChatGPT} \\
        \midrule
        
        \textbf{Total Tests} & 1403 & 1482 & 1425 & 964 & 949 & 970 \\
        
        \textbf{Syntactically Correct (Original)} & 100.00\% & 100.00\% & 100.00\% & 100.00\% & 100.00\% & 100.00\% \\
        \textbf{Syntactically Correct (Repaired)} & 100.00\% & 100.00\% & 100.00\% & 100.00\% & 100.00\% & 100.00\% \\
        
        \textbf{Compilable (Original)} & 100.00\% & 100.00\% & 100.00\% & 98.99\% & 97.98\% & 98.99\% \\
        \textbf{Compilable (Repaired)} & 100.00\% & 100.00\% & 100.00\% & 100.00\% & 100.00\% & 100.00\% \\
        
        \textbf{Passed (Original)} & 0.57\% & 6.07\% & 0.56\% & 20.02\% & 51.53\% & 18.66\% \\
        \textbf{Passed (Repaired)} & 81.18\% & 76.11\% & 76.49\% & 84.34\% & 82.19\% & 84.33\% \\
        \bottomrule
    \end{tabular}
    }
    \caption{Comparison of functional correctness metrics}
    \label{tab:rq1_1_correctness_metrics}
\end{table}
\subsection{Effect of Test Examples' Sources (RQ1)}

To answer RQ1, we evaluate the code quality of LLM-generated test cases using few-shot examples from three distinct sources: (1) human-written tests (created by developers), (2) SBST-generated tests (produced by Pynguin), and (3) LLM-generated tests (using GPT-4o with zero-shot prompting). Below, we explain the results obtained for each metric we used to assess the quality of the LLM-generated unit test (see Section~\ref{subsec:unit-test_quality_metrics}).

\begin{figure}[t]
    \centering
    \textbf{(a)} HumanEval
    \includegraphics[width=0.48\textwidth]{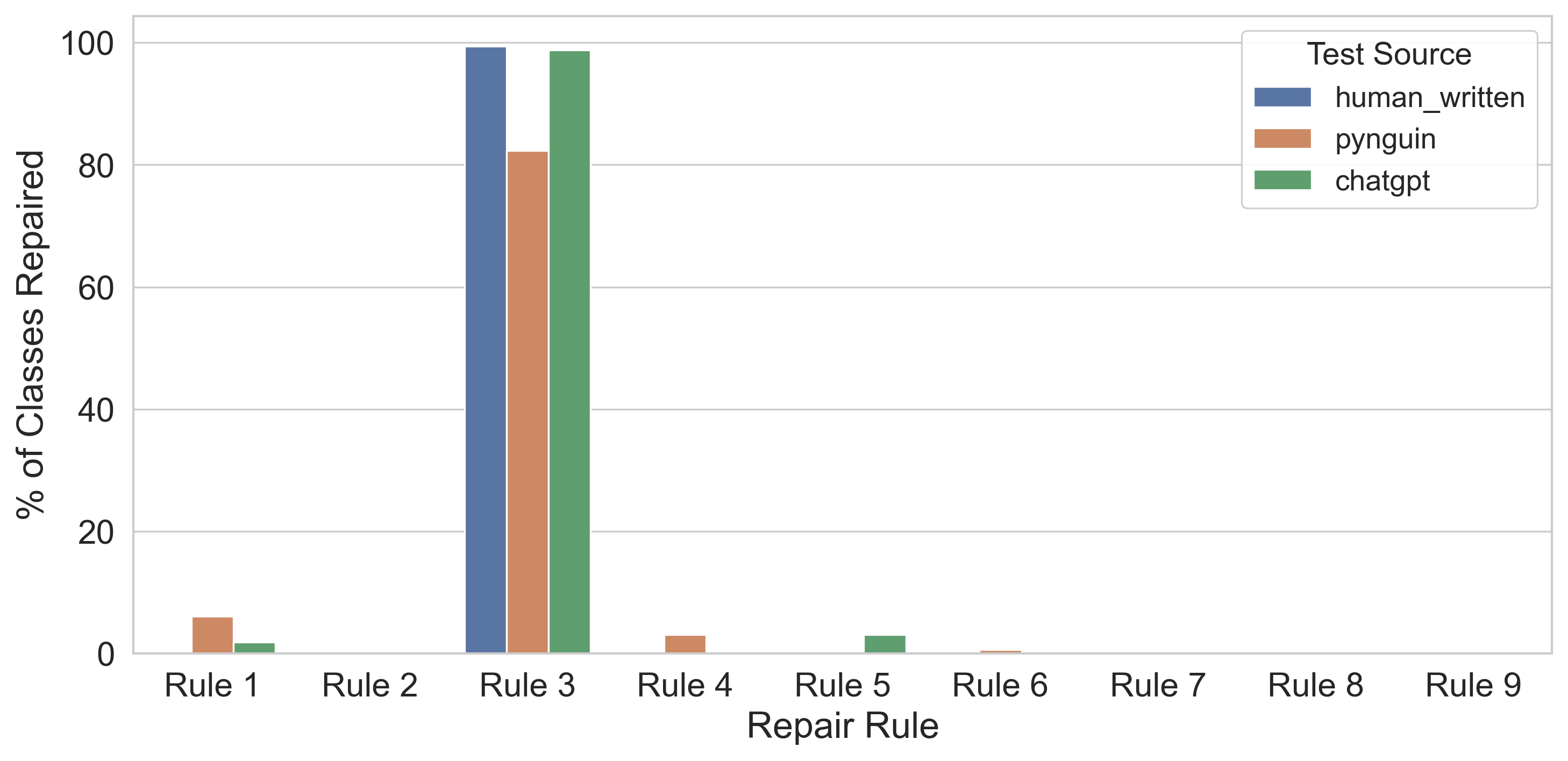}

    \par\vspace{0.15em}

    \vspace{0.4em}
     \textbf{(b)} ClassEval
    \includegraphics[width=0.48\textwidth]{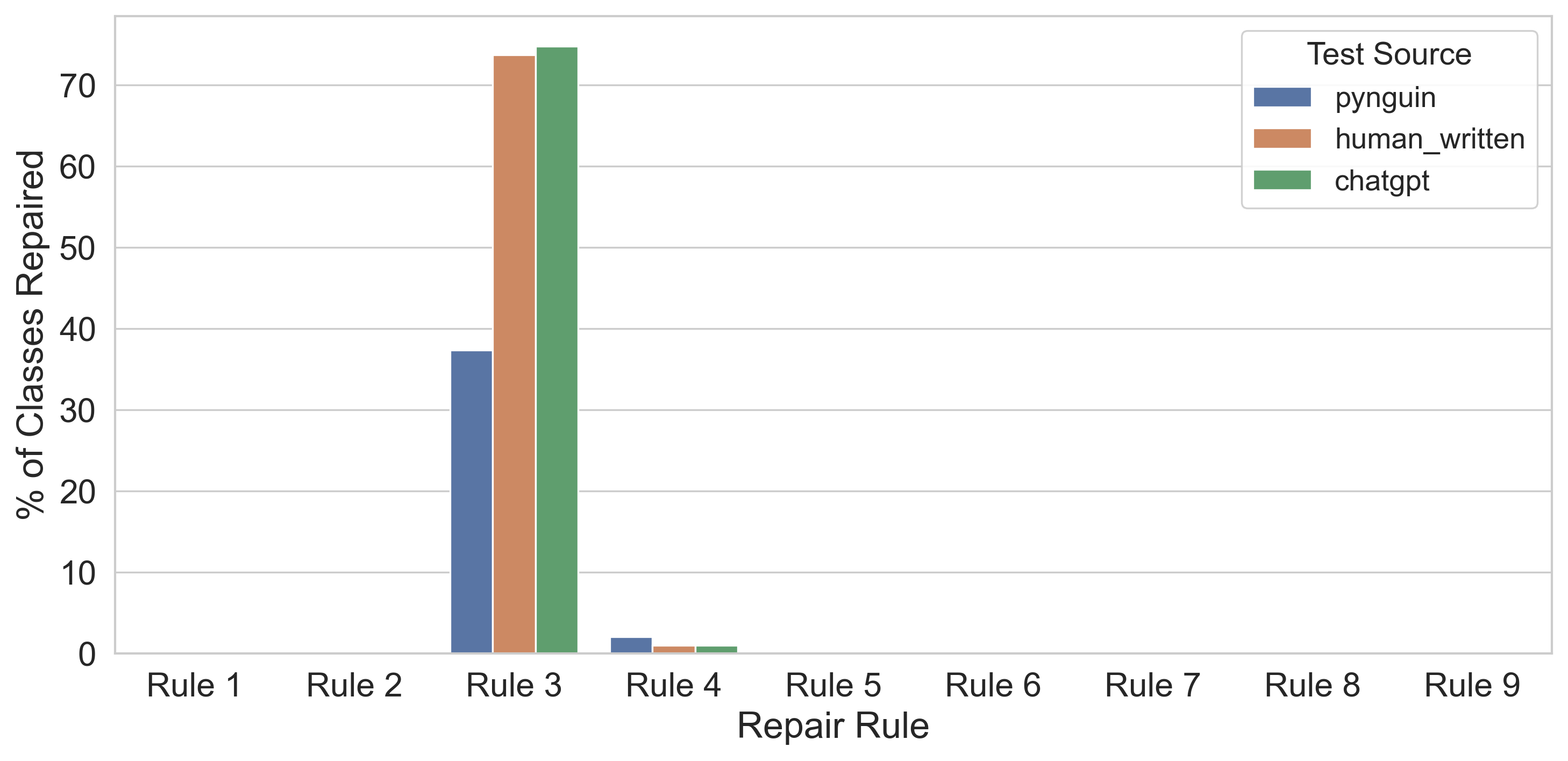}

   \caption{The distribution of the repair rules applied to LLM-generated (i.e., GPT-4.o-generated) unit tests. (a) HumanEval and (b) ClassEval results using the baseline technique for selecting unit test examples for the few-shot prompts (i.e., Random Choice from the Test Suite). The remaining graphs are available in the replication package~\cite{anonymous_2025_15561007}.
   }
    \label{fig:RQ1-rule-repair}
\end{figure}

\noindent \textbf{Functional Correctness.} As illustrated in Table~\ref{tab:rq1_1_correctness_metrics}, each approach created approximately 1,400 and 950 test cases for HumanEval and ClassEval, respectively. The results show that while all sources yield syntactically correct and mostly compilable tests, their initial functional correctness varies significantly. SBST-generated examples outperform the other test sources in generating initially correct tests, with 6.07\% pass rate for HumanEval and 51.53\% for ClassEval. This could be a result of their alignment with traditional test-generation heuristics. However, after applying the rule-based repair, the differences diminish, with all sources achieving high pass rates for both benchmarks. This indicated that LLMs mostly produce syntactically valid tests. For instance, as shown in Figure~\ref{fig:RQ1-rule-repair}.a, for the HumanEval  dataset for the baseline condition to retrieve test examples (i.e., ``Random Choice from Test Suites" technique), 6.1\% of repairs for human-written and 1.8\% for LLM-generated test cases involve Rule 1 (the elimination of incomplete test cases). Similarly, as shown in Figure~\ref{fig:RQ1-rule-repair}.b, for the ClassEval dataset for the baseline condition to retrieve test examples (i.e., ``Random Choice from Test Suites" technique), 2\% of repairs for SBST-generated tests and 1\% for LLM-generated and human-written tests involve Rule 4 (removing faulty module import statement).

Rule 3 (add missing imports to the class under test) was the most frequently applied fix, accounting for the majority of repairs. This implies that the model often overlooks critical dependencies, resulting in execution issues. However, it is important to note that importing the class under test is not always necessary in the test class, as it depends on the structure of the testing environment. 

Our findings demonstrate that LLMs can generate functional tests, although their output frequently requires lightweight post-processing to ensure functional correctness. The effectiveness of simple rule-based repair indicates that many failures stem from minor oversights rather than fundamental flaws.

\begin{table}[t]
    \centering
    \resizebox{\columnwidth}{!}{
    \begin{tabular}{lrrrrrr}
        \toprule
        & \multicolumn{3}{c}{\textbf{HumanEval}} & \multicolumn{3}{c}{\textbf{ClassEval}} \\
        \cmidrule(lr){2-4} \cmidrule(lr){5-7}
        & \textbf{ChatGPT} & \textbf{Human} & \textbf{Pynguin} & \textbf{ChatGPT} & \textbf{Human} & \textbf{Pynguin} \\
        \midrule
        \textbf{Branch Coverage} & 85.7\% & 94.0\% & 91.2\% & 82.0\% & 82.5\% & 80.0\% \\
        \textbf{Line Coverage} & 97.8\% & 98.8\% & 98.8\% & 95.5\% & 95.3\% & 95.0\% \\
        \bottomrule
    \end{tabular}
    }
    \caption{Comparison of \textit{coverage} metrics}
    \label{tab:rq1_2_coverage}
\end{table}

\noindent \textbf{Code coverage.} We measured the line and branch coverage achieved by test suites strictly consisting of LLM-generated test cases, with different sources of test examples used in few-shot prompting. The results in Table \ref{tab:rq1_2_coverage} indicate that human-written examples consistently yield the highest or comparable coverage metrics for both datasets, suggesting that they provide better patterns for the LLM to learn from. However, the test suites generated using SBST-based and LLM-generated examples also achieved high coverage, demonstrating their suitability for automated few-shot LLM test generation.

\begin{table}[t]
    \centering
    \resizebox{\columnwidth}{!}{
    \begin{tabular}{lrrrrrr}
        \toprule
        & \multicolumn{3}{c}{\textbf{HumanEval}} & \multicolumn{3}{c}{\textbf{ClassEval}} \\
        \cmidrule(lr){2-4} \cmidrule(lr){5-7}
        \textbf{Metric} & \textbf{ChatGPT} & \textbf{Human} & \textbf{Pynguin} & \textbf{ChatGPT} & \textbf{Human} & \textbf{Pynguin} \\
        \midrule
        \textbf{No. of tests} & 1,096 & 1,206 & 1,114 & 818 & 813 & 780 \\
                       
        \textbf{Cyclomatic Complexity} & 1,090 & 1,140 & 1,053 & 888 & 877 & 841 \\
        \textbf{Cognitive Complexity} & 0 & 3 & 1 & 8 & 11 & 7 \\
        \textbf{Avg. Technical Debt/Squale} & 0.18 & 0.33 & 0.25 & 0.55 & 0.52 & 0.57 \\
        \textbf{Avg. Code Smells} & 0.02 & 0.03 & 0.03 & 0.06 & 0.05 & 0.05 \\
        \bottomrule
    \end{tabular}
    }
    \caption{Comparison of \textit{code quality} metrics}
    \label{tab:rq1_3_complexity_metrics}
\end{table}

\begin{table*}[htbp]
    \centering
    \scriptsize
    \resizebox{\textwidth}{!}{%
    \begin{tabular}{@{}lrrrrrrrrrrrrrrrrrr@{}}
        \toprule
        & \multicolumn{6}{c}{\textbf{ChatGPT}} 
        & \multicolumn{6}{c}{\textbf{Human Written}} 
        & \multicolumn{6}{c}{\textbf{Pynguin}} \\
        \cmidrule(lr){2-7} \cmidrule(lr){8-13} \cmidrule(lr){14-19}
        \textbf{Selection Mode} 
        & \rotatebox{90}{Code} 
        & \rotatebox{90}{Code+Comment} 
        & \rotatebox{90}{Problem+Code} 
        & \rotatebox{90}{Problem} 
        & \rotatebox{90}{Random From All} 
        & \rotatebox{90}{Random From Class} 
        & \rotatebox{90}{Code} 
        & \rotatebox{90}{Code+Comment} 
        & \rotatebox{90}{Problem+Code} 
        & \rotatebox{90}{Problem} 
        & \rotatebox{90}{Random From All} 
        & \rotatebox{90}{Random From Class} 
        & \rotatebox{90}{Code} 
        & \rotatebox{90}{Code+Comment} 
        & \rotatebox{90}{Problem+Code} 
        & \rotatebox{90}{Problem} 
        & \rotatebox{90}{Random From All} 
        & \rotatebox{90}{Random From Class} \\
        \midrule

        \multicolumn{19}{l}{\textbf{HumanEval}} \\
        \quad Total Tests 
        & 1538 & 1515 & 1561 & 1494 & 1425 & 1050 
        & 1480 & 1488 & 1471 & 1468 & 1403 & 1588 
        & 1510 & 1524 & 1557 & 1487 & 1482 & 1679 \\
        
        \multicolumn{19}{l}{Syntactically Correct (\%)} \\
        \quad\quad Before Repair 
        & 100 & 100 & 100 & 100 & 100 & 99.4 
        & 99.4 & 100 & 100 & 100 & 100 & 100 
        & 100 & 99.4 & 100 & 100 & 100 & 100 \\
        \quad\quad After Repair 
        & 100 & 100 & 100 & 100 & 100 & 100 
        & 100 & 100 & 100 & 100 & 100 & 100 
        & 100 & 100 & 100 & 100 & 100 & 100 \\

        \multicolumn{19}{l}{Compilable (\%)} \\
        \quad\quad Before Repair 
        & 100 & 100 & 100 & 100 & 100 & 100 
        & 100 & 100 & 100 & 100 & 100 & 100 
        & 98.2 & 100 & 100 & 98.8 & 100 & 90.2 \\
        \quad\quad After Repair 
        & 100 & 100 & 100 & 100 & 100 & 100 
        & 100 & 100 & 100 & 100 & 100 & 100 
        & 100 & 100 & 100 & 100 & 100 & 100 \\

        \quad Passed (\%) \\
        \quad\quad Before Repair 
        & 0 & 0 & 0 & 0.6 & 0.6 & 0 
        & 0 & 0.5 & 0.3 & 0 & 0.6 & 0 
        & 7.8 & 6.4 & 3.9 & 5.9 & 6.1 & 16.0 \\
        \quad\quad After Repair 
        & 79.0 & 77.4 & 79.3 & 78.8 & 76.5 & 73.3 
        & 80.5 & 78.7 & 79.4 & 79.3 & 81.2 & 78.7 
        & 74.3 & 75.7 & 74.7 & 73.0 & 76.1 & 71.6 \\

        \addlinespace

        \multicolumn{19}{l}{\textbf{ClassEval}} \\
        \quad Total Tests 
        & 965 & 988 & 977 & 988 & 970 & 873 
        & 979 & 972 & 993 & 988 & 964 & 1073 
        & 960 & 987 & 945 & 937 & 949 & 1029 \\
        
        \multicolumn{19}{l}{Syntactically Correct (\%)} \\
        \quad\quad Before Repair 
        & 98.99 & 100 & 100 & 98.99 & 100 & 100 
        & 98.99 & 98.99 & 100 & 100 & 100 & 100 
        & 100 & 97.98 & 98.99 & 98.99 & 100 & 98.99 \\
        \quad\quad After Repair 
        & 100 & 100 & 100 & 100 & 100 & 100 
        & 100 & 100 & 100 & 100 & 100 & 100 
        & 100 & 100 & 100 & 100 & 100 & 98.99 \\

        \multicolumn{19}{l}{Compilable (\%)} \\
        \quad\quad Before Repair 
        & 97.98 & 98.99 & 97.98 & 97.98 & 98.99 & 98.99 
        & 98.99 & 97.98 & 97.98 & 97.98 & 98.99 & 100 
        & 100 & 98.99 & 100 & 98.99 & 97.98 & 98.99 \\
        \quad\quad After Repair 
        & 100 & 100 & 100 & 100 & 100 & 100 
        & 100 & 100 & 100 & 100 & 100 & 100 
        & 100 & 100 & 100 & 100 & 100 & 100 \\

        \quad Passed (\%) \\
        \quad\quad Before Repair 
        & 17.72 & 19.13 & 17.91 & 13.66 & 18.66 & 4.58 
        & 21.45 & 26.13 & 18.63 & 19.64 & 20.02 & 7.92 
        & 51.04 & 50.25 & 61.27 & 57.84 & 51.53 & 57.53 \\
        \quad\quad After Repair 
        & 83.42 & 83.10 & 84.24 & 84.82 & 84.33 & 76.17 
        & 82.33 & 83.95 & 84.59 & 84.01 & 84.34 & 79.12 
        & 83.12 & 83.38 & 84.02 & 83.14 & 82.19 & 81.44 \\
        \bottomrule
    \end{tabular}%
    }
    \caption{Comparison of functional correctness metrics across different sources and selection modes}
    \label{tab:rq2_1_functional_correctness}
\end{table*}

\begin{table*}[htbp]
    \centering
    \scriptsize
    \resizebox{\textwidth}{!}{
    \begin{tabular}{lrrrrrrrrrrrrrrrrrr}
        \toprule
        & \multicolumn{6}{c}{\textbf{ChatGPT}} & \multicolumn{6}{c}{\textbf{Human Written}} & \multicolumn{6}{c}{\textbf{Pynguin}} \\
        \cmidrule(lr){2-7} \cmidrule(lr){8-13} \cmidrule(lr){14-19}
        \textbf{Selection Mode} & \rotatebox{90}{Code} & \rotatebox{90}{Code+Comment} & \rotatebox{90}{Problem+Code} & \rotatebox{90}{Problem} & \rotatebox{90}{Random From All} & \rotatebox{90}{Random From Class} & \rotatebox{90}{Code} & \rotatebox{90}{Code+Comment} & \rotatebox{90}{Problem+Code} & \rotatebox{90}{Problem} & \rotatebox{90}{Random From All} & \rotatebox{90}{Random From Class} & \rotatebox{90}{Code} & \rotatebox{90}{Code+Comment} & \rotatebox{90}{Problem+Code} & \rotatebox{90}{Problem} & \rotatebox{90}{Random From All} & \rotatebox{90}{Random From Class} \\
        \midrule
        
        \multicolumn{19}{l}{\textbf{HumanEval}} \\
        \quad Branch Coverage & 93.5\% & 93.3\% & 92.5\% & 94.6\% & 92.2\% & 59.4\% & 96.9\% & 96.7\% & 96.3\% & 95.3\% & 96.1\% & 93.5\% & 91.0\% & 92.0\% & 93.6\% & 97.1\% & 88.1\% & 90.1\% \\
        \quad Line Coverage & 99.0\% & 98.9\% & 98.4\% & 99.0\% & 98.8\% & 88.6\% & 99.4\% & 99.6\% & 99.3\% & 99.1\% & 99.2\% & 99.1\% & 98.6\% & 98.7\% & 99.1\% & 99.5\% & 98.1\% & 98.6\% \\ 
        \addlinespace
        
        \multicolumn{19}{l}{\textbf{ClassEval}} \\
        \quad Branch Coverage & 79.6\% & 80.7\% & 83.6\% & 82.0\% & 82.0\% & 68.2\% & 82.3\% & 81.3\% & 85.3\% & 83.1\% & 82.5\% & 73.9\% & 82.3\% & 81.8\% & 82.0\% & 77.2\% & 80.0\% & 81.7\% \\
        \quad Line Coverage & 94.8\% & 95.4\% & 95.7\% & 95.6\% & 95.5\% & 88.4\% & 94.9\% & 94.8\% & 96.8\% & 95.8\% & 95.3\% & 94.0\% & 95.3\% & 96.1\% & 95.2\% & 93.4\% & 95.0\% & 95.8\% \\
        
        \bottomrule
    \end{tabular}
    }
    \caption{Comparison of coverage metrics across different test sources and selection modes}
    \label{tab:rq2_2_coverage_metrics}
\end{table*}

\noindent \textbf{Code quality.} As displayed in Table \ref{tab:rq1_3_complexity_metrics}, cyclomatic complexity appears strongly correlated with test suite size, varying with the number of test cases included. On the other hand, cognitive complexity shows greater dependence on the complexity of the benchmark, with consistently higher values for ClassEval across all test sources. Test suites generated using human-written test examples exhibited higher cognitive complexity than other test sources, suggesting they may produce more complex test logic. Technical debt and code smell showed minimal differences over the test sources for both HumanEval and ClassEval datasets. The data further indicates that, with the higher complexity of the benchmark, both average technical debt and the number of code smells increase for all test sources.

\subsection{Effect of Example Selection Techniques (RQ2)}

The results obtained regarding how different retrieval-based techniques affect the quality of the LLM-generated tests are as follows:

\noindent \textbf{Functional correctness.} As shown in Table  \ref{tab:rq2_1_functional_correctness}, all selection methods show poor initial correctness, significantly improving after repair. The rule-based selection methods yield comparable values to the \textit{Random Choice from Test Suite} (``Random from All") baseline, which selects random test classes from the test suite. The \textit{Problem and source code similarity} (``Problem+Code") selection technique, which combines source code and problem similarity, yields consistently high coverage. Still, the baseline method is not particularly behind. Hence, we conclude that the selection techniques have no significant effect on the correctness of the generated tests.

\noindent \textbf{Code Coverage.} 
Results show that test example selection methods still have a considerable impact on coverage. \textit{Problem and source code similarity} ("Problem+Code") selection technique is the most promising among the retrieval-based methods. It achieves the highest branch coverage on the ClassEval dataset across LLM-based (GPT-4.o) and human-written test sources. For SBST-based test sources (i.e., Pynguin), ``Problem+Code" retrieval technique achieves branch coverage that is comparable with the Problem similarity retrieval technique. The obtained results  illustrate potential of the ``Problem+Code" retrieval technique in selecting contextually relevant test examples to guide the LLM in generating diverse test cases. Other selection techniques display on par performance relative to the baseline. Conversely, the \textit{Random Choice from Class under Test} ("Random from Class") technique consistently underperforms, suggesting that providing the demonstrations solely from the class under test is ineffective.

\noindent \textbf{Code quality}. The evaluation of different retrieval-based methods does not reveal any distinct patterns in their impact on the code quality of the tests. None of the retrieval-based techniques displays a consistent advantage or disadvantage over the baseline. The results show substantial variation depending on both the measured dimension and the evaluated dataset. While some methods, such as \textit{Problem and source code similarity}, demonstrate better maintainability scores compared to the baseline in certain scenarios, these advantages are not consistently sustained.

\begin{table}[t]
    \centering
    \resizebox{\columnwidth}{!}{
    \begin{tabular}{lrrrrrr}
        \toprule
        & \multicolumn{3}{c}{\textbf{HumanEval}} & \multicolumn{3}{c}{\textbf{ClassEval}} \\
        \cmidrule(lr){2-4} \cmidrule(lr){5-7}
        \textbf{Metric} & \textbf{ChatGPT} & \textbf{Human} & \textbf{Pynguin} & \textbf{ChatGPT} & \textbf{Human} & \textbf{Pynguin} \\
        \midrule
        \textbf{Total Tests} & 1,096 & 1,206 & 987 & 819 & 814 & 780 \\
        \textbf{Kept Tests (\%)} & 1.73\% & 1.33\% & 10.33\% & 4.88\% & 2.46\% & 23.33\% \\
        \textbf{Removed Tests (\%)} & 7.39\% & 5.47\% & 9.83\% & 40.17\% & 5.65\% & 15.13\% \\
        \textbf{Skipped Tests (\%)} & 90.88\% & 93.20\% & 77.41\% & 36.26\% & 71.62\% & 9.10\% \\
        \textbf{Faulty Tests (\%)} & 0\% & 0\% & 2.43\% & 18.68\% & 20.27\% & 52.44\% \\
        \bottomrule
    \end{tabular}
    }
    \caption{Test suite optimisation results showing percentages of kept (coverage-improving), removed, skipped, and failed tests}
    \label{tab:rq1_4_optimised_test_counts}
\end{table}

\begin{table}[t]
    \centering
    \resizebox{\columnwidth}{!}{
    \begin{tabular}{lrrrrrr}
        \toprule
        & \multicolumn{3}{c}{\textbf{HumanEval}} & \multicolumn{3}{c}{\textbf{ClassEval}} \\
        \cmidrule(lr){2-4} \cmidrule(lr){5-7}
        \textbf{Metric} & \textbf{ChatGPT} & \textbf{Human} & \textbf{Pynguin} & \textbf{ChatGPT} & \textbf{Human} & \textbf{Pynguin} \\
        \midrule
        
        \multicolumn{7}{l}{\textbf{Branch Coverage}} \\
        \quad Initial & 92.8\% & 95.5\%  & 64.5\% & 76.2\% & 93.1\% & 36.8\% \\
        \quad Optimised & 98.4\% & 98.2\% &   91.4\% & 82.4\% & 93.8\% & 60.7\% \\
        \quad $\Delta$ & +5.6\% & +2.7\% & +26.9\% & +6.2\% & +0.7\% & +23.9\% \\
        \addlinespace
        
        \multicolumn{7}{l}{\textbf{Line Coverage}} \\
        \quad Initial & 98.6\% & 99.0\% & 89.0\% & 93.2\% & 97.4\% & 68.2\% \\
        \quad Optimised & 99.7\% & 99.7\%  & 97.8\% & 95.2\% & 97.5\% & 83.1\% \\
        \quad $\Delta$ & +1.1\% & +0.7\% & +8.8\% & +2.0\% & +0.1\% & +14.9\% \\
        
        \bottomrule
    \end{tabular}
    }
    \caption{Comparison of coverage metrics before (Initial) and after optimisation (Optimised)}
    \label{tab:rq1_4_optimised_coverage_metrics}
\end{table}

\begin{table}[t]
    \centering
    \resizebox{\columnwidth}{!}{
    \begin{tabular}{lrrrrrr}
        \toprule
        & \multicolumn{3}{c}{\textbf{HumanEval}} & \multicolumn{3}{c}{\textbf{ClassEval}} \\
        \cmidrule(lr){2-4} \cmidrule(lr){5-7}
        \textbf{Metric} & \textbf{ChatGPT} & \textbf{Human} & \textbf{Pynguin} & \textbf{ChatGPT} & \textbf{Human} & \textbf{Pynguin} \\
        \midrule
        
        \multicolumn{7}{l}{\textbf{Cyclomatic Complexity}} \\
        \quad Initial & 969 & 1,175 & 211 & 713 & 2,318 & 229 \\
        \quad Optimised & 988 & 1,191 & 313 & 754 & 2,338 & 414 \\
        \quad $\Delta$ & +19 & +16 & +102 & +41 & +20 & +185 \\
        \addlinespace
        
        \multicolumn{7}{l}{\textbf{Cognitive Complexity}} \\
        \quad Initial & 0 & 0 & 2 & 9 & 215 & 0 \\
        \quad Optimised & 0 & 0 & 2 & 10 & 215 & 3 \\
        \quad $\Delta$ & 0 & 0 & 0 & +1 & 0 & +3 \\
        \addlinespace
        
        \multicolumn{7}{l}{\textbf{Avg Technical Debt (minutes)}} \\
        \quad Initial & 0.14 & 0.67 & 2.15 & 0.56 & 2.05 & 1.49 \\
        \quad Optimised & 0.15 & 0.66 & 1.16 & 0.53 & 5.45 & 1.38 \\
        \quad $\Delta$ & +0.01 & -0.01 & -0.99 & -0.03 & +3.40 & -0.11 \\
        \addlinespace
        
        \multicolumn{7}{l}{\textbf{Avg Code Smells}} \\
        \quad Initial & 0.02 & 0.09 & 0.36 & 0.06 & 0.30 & 0.28 \\
        \quad Optimised & 0.02 & 0.09 & 0.19 & 0.05 & 0.79 & 0.25 \\
        \quad $\Delta$ & +0.00 & -0.00 & -0.17 & -0.00 & +0.49 & -0.03 \\
        
        \bottomrule
    \end{tabular}
    }
    \caption{Comparison of code quality metrics before (Initial) and after optimisation (Optimised)}
    \label{tab:rq1_4_optimised_complexity_metrics}
\end{table}

\subsection{Example Sources for Test Suite Enhancement (RQ3)}

The experimental results presented in Tables \ref{tab:rq1_4_optimised_test_counts} and \ref{tab:rq1_4_optimised_coverage_metrics} reveal several key insights about test suite enhancement capabilities of the LLMs through few-shot learning. Analysis shows that SBST-generated (Pynguin) examples produced the highest percentage of coverage-improving tests and achieved the most significant improvement in coverage, followed by LLM-based examples, with human-written examples showing the least improvements. However, one should interpret this pattern in context: the human-created initial test suite already demonstrates superior coverage, leaving less room for improvement, whereas the poorer coverage of the SBST-based initial suite creates greater potential for enhancement.

Overall, few-shot learning from all sources of test demonstrations effectively improves the coverage of the initial test suite, with the optimized coverage approaching 100\% for all sources on HumanEval. For ClassEval, both human and LLM-based examples achieved high coverage, while the SBST-based suite lagged. Notably, the optimization phase revealed higher test error rates on the ClassEval benchmark, particularly those generated using Pynguin examples, suggesting that integrating new tests into existing suites requires a more sophisticated adaptation technique for complex codebases. As shown in Table~\ref{tab:rq1_4_optimised_complexity_metrics}, there is a negligible change in code quality  except for the cyclomatic complexity values, for which we observe a significant increase. The increase in cyclomatic complexity for ClassEval is higher than that of HumanEval. Moreover, the highest cyclomatic complexity is observed when test example sources are SBST-based (i.e., Pynguin-generated).

\subsection{Example Selection for Test Suite Enhancement (RQ4)}
On both HumanEval and ClassEval datasets, the most considerable improvements in branch and line coverage are observed for \textit{Problem definition similarity} and \textit{Problem definition similarity} retrieval techniques, although the differences are slight.  Regarding correctness, we did not observe any faulty tests in the HumanEval dataset, while in the ClassEval the highest percentage of faulty tests was observed for \textit{Source code similarity} retrieval technique across all three test example sources (i.e., developers, SBST, LLMs). Finally, on the HumanEval dataset, the largest improvement in cyclomatic complexity is observed for the \textit{Random Choice from Class under Test} retrieval technique. We did not observe any improvement in cognitive complexity, but we observed slight improvements in average technical debt and average code smells. Results for the ClassEval dataset followed similar trends.

\subsection{Threats to Validity}
\noindent{}\textbf{Construct validity.} To mitigate mono-operation bias, we employed a diverse set of metrics to evaluate various aspects of test quality referring to established approaches in test generation research \cite{wang_software_2023, ouedraogo_large-scale_2024, siddiq_using_2024}. The selected metrics evaluate LLM-generated tests—beyond correctness and coverage—by assessing readability, maintainability, and robustness, enabling meaningful comparisons across few-shot prompting
methods. However, although code coverage is widely used as an indicator of test suite effectiveness \cite{zhang_assessing_2024, ahmed_can_2016}, its correlation with actual defect detection depends on factors like test quantity \cite{inozemtseva_coverage_2014}. Therefore, as future work, one can enhance test code quality evaluation methods by incorporating richer metrics, such as mutation testing, to better capture test suite effectiveness beyond coverage alone.

\noindent \textbf{Internal validity.} We used datasets containing tests implemented by (human) developers (i.e., HumanEval and ClassEval). Since the test cases across HumanEval and ClassEval datasets vary in format, we converted test cases in these two benchmarks into \texttt{pytest} to form human-written tests. We generate SBST-based tests using Pynguin. \cite{noauthor_pynguinpython_nodate}, an established automatic unit test generation tool based on SBST. To create the dataset consisting of LLM-generated tests, we used GPT-4o using a simple prompt inspired by prior studies \cite{xie_chatunitest_2023, ouedraogo_large-scale_2024}. We could conduct manual checks only for statistically representative samples due to large number of LLM-generated unit tests. The number of examples to include in the prompts ($n_{examples}$) acts as a confounding factor (i.e., few-shot prompt performance varies depending on the number of examples). Therefore, we tested three prompt configurations to determine the optimal number of tests to include in the prompts. Based on our findings, we used five examples in the prompts. 

\noindent{}\textbf{External validity.} HumanEval is a simple code benchmark consisting of isolated coding exercises. Therefore, we also included in our study the ClassEval dataset, containing cross-file references that require tests with class-level references to achieve high coverage. In future work, expanding the benchmark to include real-world projects (e.g., SWE-bench, LiveCodeBench) across diverse domains and languages will improve the practical relevance of results. Finally, we conducted our experiments using GPT-4.o. Therefore, future research should explore a broader range of LLM architectures beyond GPT-4o to assess how model differences impact test generation.


\section{Discussion}
\label{sec:discussion}
\noindent \textbf{Comparison with Prior Work.} Prior work shows that LLMs can generate readable, high-coverage tests, but correctness and usability remain limited without validation or repair (e.g., Sch\"{a}fer et al.~\cite{schafer_adaptive_2023}, Siddiq et al.~\cite{siddiq_fault_2024}, Xie et al.~\cite{xie_chatunitest_2023}). Our results reinforce this finding: few-shot prompting alone rarely produces executable tests, yet many failures arise from systematic, lightweight issues that can be addressed with simple rule-based repair.

Our findings also align with Ou\'{e}draogo et al.~\cite{ouedraogo_large-scale_2024}, who show that few-shot prompting yields only modest gains in correctness and coverage over zero-shot prompting, while improving readability and reducing syntactic errors. We similarly observe structural improvements without resolving fundamental correctness issues. We extend their work by systematically evaluating automated, retrieval-based selection of few-shot examples from heterogeneous sources, showing that example provenance and selection strategy affect coverage and downstream utility, even if correctness gains remain limited.

More broadly, our results refine prior few-shot and hybrid approaches (e.g., Alshahwan et al.~\cite{alshahwan_assured_2024, alshahwan_automated_2024}, CODAMOSA~\cite{lemieux_codamosa_2023}), which show that LLM-generated tests are most effective when embedded in pipelines that filter, validate, or augment existing tests. We confirm that few-shot prompting is most valuable for enhancing existing test suites—especially when initial coverage is low—rather than for generating standalone suites, a contrast to much of the earlier literature. Our work complements studies by Yang et al.~\cite{yang_evaluation_2024} and Siddiq et al.~\cite{siddiq_using_2024, siddiq_fault_2024} on prompt design and context inclusion by shifting focus from what information to include to \textit{which tests to use as demonstrations} and \textit{how to select them}. Notably, we challenge the assumption that human-written tests are always optimal: while they yield strong standalone coverage, SBST-generated tests can be equally effective or superior for test suite enhancement.

Finally, unlike tool-centric efforts such as ChatUniTest~\cite{xie_chatunitest_2023} or EvalPlus~\cite{liu_is_2023}, we do not propose a new generation or repair framework. Instead, we provide an empirical synthesis of how few-shot prompting interacts with example source, selection strategy, and test suite context. Therefore, we reframe LLM-based test generation as a program comprehension and maintenance aid that integrates with, rather than replaces, human-written tests.\\
\noindent \textbf{Implications for Researchers.} Our results highlight the importance of evaluating LLM-generated tests beyond correctness and coverage alone. Maintainability-related properties, such as cognitive complexity, test smells, and technical debt are essential for understanding how generated tests affect program comprehension and long-term evolution. Future research should therefore treat test code quality as a first-class evaluation dimension. Moreover, while retrieval-based example selection can improve coverage, its benefits are not uniform across all quality metrics. This suggests opportunities for future work on adaptive or multi-objective selection strategies that balance relevance, diversity, and maintainability. Finally, the observed differences between HumanEval and ClassEval reinforce the need for diverse and realistic benchmarks when evaluating LLM-based test generation.\\
\noindent \textbf{Implications for Practitioners.} For practitioners, our findings suggest that LLM-generated tests are most effective when used as augmentation mechanisms rather than replacements for human-written test suites. Few-shot prompting guided by existing tests can help fill coverage gaps, particularly in under-tested systems, but generated tests should be validated, repaired, and filtered before integration. The effectiveness of lightweight rule-based repair indicates that adopting LLM-based test generation does not require complex infrastructure. At the same time, the comparable maintainability of LLM-generated tests observed in our study suggests that, when properly guided, such tests need not impose excessive technical debt, supporting their practical viability in hybrid human–AI development workflows.
\section{Conclusions}
\label{sec:conclusion}
Our evaluation shows that LLMs can reliably generate syntactically valid tests using few-shot prompting, with human-written demonstrations yielding the highest post-repair correctness and coverage. While SBST examples produced more functional tests before repair, they underperformed in readability. For test suite enhancement, SBST examples improved low-coverage suites the most, but human-written tests ultimately achieved the highest final coverage. Regarding RQ2, retrieval-based example selection had a limited impact overall. However, the problem and source code similarity method consistently produced high-quality, high-coverage tests and showed the most promise. Overall, these results highlight the effectiveness of LLMs in generating functional, high-coverage unit tests and guide for improving prompt design and test generation.

\bibliographystyle{ACM-Reference-Format}
\bibliography{sample-base}


\begin{thebibliography}{46}


\ifx \showCODEN    \undefined \def \showCODEN     #1{\unskip}     \fi
\ifx \showISBNx    \undefined \def \showISBNx     #1{\unskip}     \fi
\ifx \showISBNxiii \undefined \def \showISBNxiii  #1{\unskip}     \fi
\ifx \showISSN     \undefined \def \showISSN      #1{\unskip}     \fi
\ifx \showLCCN     \undefined \def \showLCCN      #1{\unskip}     \fi
\ifx \shownote     \undefined \def \shownote      #1{#1}          \fi
\ifx \showarticletitle \undefined \def \showarticletitle #1{#1}   \fi
\ifx \showURL      \undefined \def \showURL       {\relax}        \fi
\providecommand\bibfield[2]{#2}
\providecommand\bibinfo[2]{#2}
\providecommand\natexlab[1]{#1}
\providecommand\showeprint[2][]{arXiv:#2}

\bibitem[noa({[n.\,d.]})]%
        {noauthor_pynguinpython_nodate}
 \bibinfo{year}{[n.\,d.]}\natexlab{}.
\newblock \bibinfo{title}{Pynguin—{PYthoN} {General} {UnIt} test {geNerator}
  — pynguin 0.41.0.dev documentation}.
\newblock
\urldef\tempurl%
\url{https://pynguin.readthedocs.io/en/latest/}
\showURL{%
\tempurl}
\newblock
\shownote{Accessed: 2025-01-27}.


\bibitem[Achiam et~al\mbox{.}(2023)]%
        {openai_gpt-4_2023}
\bibfield{author}{\bibinfo{person}{OpenAI~Josh Achiam}, \bibinfo{person}{Steven
  Adler}, \bibinfo{person}{Sandhini Agarwal}, \bibinfo{person}{Lama Ahmad},
  \bibinfo{person}{Ilge Akkaya}, \bibinfo{person}{Florencia~Leoni Aleman},
  \bibinfo{person}{Diogo Almeida}, \bibinfo{person}{Janko Altenschmidt},
  \bibinfo{person}{Sam Altman}, \bibinfo{person}{Shyamal Anadkat},
  \bibinfo{person}{Red Avila}, \bibinfo{person}{Igor Babuschkin},
  \bibinfo{person}{Suchir Balaji}, \bibinfo{person}{Valerie Balcom},
  \bibinfo{person}{Paul Baltescu}, \bibinfo{person}{Haim ing Bao},
  \bibinfo{person}{Mo Bavarian}, \bibinfo{person}{Jeff Belgum},
  \bibinfo{person}{Irwan Bello}, \bibinfo{person}{Jake Berdine},
  \bibinfo{person}{Gabriel Bernadett-Shapiro}, \bibinfo{person}{Christopher
  Berner}, \bibinfo{person}{Lenny Bogdonoff}, \bibinfo{person}{Oleg Boiko},
  \bibinfo{person}{Made laine Boyd}, \bibinfo{person}{Anna-Luisa Brakman},
  \bibinfo{person}{Greg Brockman}, \bibinfo{person}{Tim Brooks},
  \bibinfo{person}{Miles Brundage}, \bibinfo{person}{Kevin Button},
  \bibinfo{person}{Trevor Cai}, \bibinfo{person}{Rosie Campbell},
  \bibinfo{person}{Andrew Cann}, \bibinfo{person}{Brittany Carey},
  \bibinfo{person}{Chelsea Carlson}, \bibinfo{person}{Rory Carmichael},
  \bibinfo{person}{Brooke Chan}, \bibinfo{person}{Che Chang},
  \bibinfo{person}{Fotis Chantzis}, \bibinfo{person}{Derek Chen},
  \bibinfo{person}{Sully Chen}, \bibinfo{person}{Ruby Chen},
  \bibinfo{person}{Jason Chen}, \bibinfo{person}{Mark Chen},
  \bibinfo{person}{Benjamin Chess}, \bibinfo{person}{Chester Cho},
  \bibinfo{person}{Casey Chu}, \bibinfo{person}{Hyung~Won Chung},
  \bibinfo{person}{Dave Cummings}, \bibinfo{person}{Jeremiah Currier},
  \bibinfo{person}{Yunxing Dai}, \bibinfo{person}{Cory Decareaux},
  \bibinfo{person}{Thomas Degry}, \bibinfo{person}{Noah Deutsch},
  \bibinfo{person}{Damien Deville}, \bibinfo{person}{Arka Dhar},
  \bibinfo{person}{David Dohan}, \bibinfo{person}{Steve Dowling},
  \bibinfo{person}{Sheila Dunning}, \bibinfo{person}{Adrien Ecoffet},
  \bibinfo{person}{Atty Eleti}, \bibinfo{person}{Tyna Eloundou},
  \bibinfo{person}{David Farhi}, \bibinfo{person}{Liam Fedus},
  \bibinfo{person}{Niko Felix}, \bibinfo{person}{Sim'on~Posada Fishman},
  \bibinfo{person}{Juston Forte}, \bibinfo{person}{Is abella Fulford},
  \bibinfo{person}{Leo Gao}, \bibinfo{person}{Elie Georges},
  \bibinfo{person}{Christian Gibson}, \bibinfo{person}{Vik Goel},
  \bibinfo{person}{Tarun Gogineni}, \bibinfo{person}{Gabriel Goh},
  \bibinfo{person}{Raphael Gontijo-Lopes}, \bibinfo{person}{Jonathan Gordon},
  \bibinfo{person}{Morgan Grafstein}, \bibinfo{person}{Scott Gray},
  \bibinfo{person}{Ryan Greene}, \bibinfo{person}{Joshua Gross},
  \bibinfo{person}{Shixiang~Shane Gu}, \bibinfo{person}{Yufei Guo},
  \bibinfo{person}{Chris Hallacy}, \bibinfo{person}{Jesse Han},
  \bibinfo{person}{Jeff Harris}, \bibinfo{person}{Yuchen He},
  \bibinfo{person}{Mike Heaton}, \bibinfo{person}{Jo hannes Heidecke},
  \bibinfo{person}{Chris Hesse}, \bibinfo{person}{Alan Hickey},
  \bibinfo{person}{Wade Hickey}, \bibinfo{person}{Peter Hoeschele},
  \bibinfo{person}{Brandon Houghton}, \bibinfo{person}{Kenny Hsu},
  \bibinfo{person}{Shengli Hu}, \bibinfo{person}{Xin Hu},
  \bibinfo{person}{Joost Huizinga}, \bibinfo{person}{Shantanu Jain},
  \bibinfo{person}{Shawn Jain}, \bibinfo{person}{Joanne Jang},
  \bibinfo{person}{Angela Jiang}, \bibinfo{person}{Roger Jiang},
  \bibinfo{person}{Haozhun Jin}, \bibinfo{person}{Denny Jin},
  \bibinfo{person}{Shino Jomoto}, \bibinfo{person}{Billie Jonn},
  \bibinfo{person}{Heewoo Jun}, \bibinfo{person}{Tomer Kaftan},
  \bibinfo{person}{Lukasz Kaiser}, \bibinfo{person}{Ali Kamali},
  \bibinfo{person}{Ingmar Kanitscheider}, \bibinfo{person}{Nitish~Shirish
  Keskar}, \bibinfo{person}{Tabarak Khan}, \bibinfo{person}{Logan Kilpatrick},
  \bibinfo{person}{Jong~Wook Kim}, \bibinfo{person}{Christina Kim},
  \bibinfo{person}{Yongjik Kim}, \bibinfo{person}{Hendrik Kirchner},
  \bibinfo{person}{Jamie~Ryan Kiros}, \bibinfo{person}{Matthew Knight},
  \bibinfo{person}{Daniel Kokotajlo}, \bibinfo{person}{Lukasz Kondraciuk},
  \bibinfo{person}{Andrew Kondrich}, \bibinfo{person}{Aris Konstantinidis},
  \bibinfo{person}{Kyle Kosic}, \bibinfo{person}{Gretchen Krueger},
  \bibinfo{person}{Vishal Kuo}, \bibinfo{person}{Michael Lampe},
  \bibinfo{person}{Ikai Lan}, \bibinfo{person}{Teddy Lee}, \bibinfo{person}{Jan
  Leike}, \bibinfo{person}{Jade Leung}, \bibinfo{person}{Daniel Levy},
  \bibinfo{person}{Chak Li}, \bibinfo{person}{Rachel Lim},
  \bibinfo{person}{Molly Lin}, \bibinfo{person}{Stephanie Lin},
  \bibinfo{person}{Ma teusz Litwin}, \bibinfo{person}{Theresa Lopez},
  \bibinfo{person}{Ryan Lowe}, \bibinfo{person}{Patricia Lue},
  \bibinfo{person}{Anna Makanju}, \bibinfo{person}{Kim Malfacini},
  \bibinfo{person}{Sam Manning}, \bibinfo{person}{Todor Markov},
  \bibinfo{person}{Yaniv Markovski}, \bibinfo{person}{Bianca Martin},
  \bibinfo{person}{Katie Mayer}, \bibinfo{person}{Andrew Mayne},
  \bibinfo{person}{Bob McGrew}, \bibinfo{person}{Scott~Mayer McKinney},
  \bibinfo{person}{Christine McLeavey}, \bibinfo{person}{Paul McMillan},
  \bibinfo{person}{Jake McNeil}, \bibinfo{person}{David Medina},
  \bibinfo{person}{Aalok Mehta}, \bibinfo{person}{Jacob Menick},
  \bibinfo{person}{Luke Metz}, \bibinfo{person}{Andrey Mishchenko},
  \bibinfo{person}{Pamela Mishkin}, \bibinfo{person}{Vinnie Monaco},
  \bibinfo{person}{Evan Morikawa}, \bibinfo{person}{Daniel~P. Mossing},
  \bibinfo{person}{Tong Mu}, \bibinfo{person}{Mira Murati},
  \bibinfo{person}{Oleg Murk}, \bibinfo{person}{David M'ely},
  \bibinfo{person}{Ashvin Nair}, \bibinfo{person}{Reiichiro Nakano},
  \bibinfo{person}{Rajeev Nayak}, \bibinfo{person}{Arvind Neelakantan},
  \bibinfo{person}{Richard Ngo}, \bibinfo{person}{Hyeonwoo Noh},
  \bibinfo{person}{Ouyang Long}, \bibinfo{person}{Cullen O'Keefe},
  \bibinfo{person}{Jakub~W. Pachocki}, \bibinfo{person}{Alex Paino},
  \bibinfo{person}{Joe Palermo}, \bibinfo{person}{Ashley Pantuliano},
  \bibinfo{person}{Giambattista Parascandolo}, \bibinfo{person}{Joel Parish},
  \bibinfo{person}{Emy Parparita}, \bibinfo{person}{Alexandre Passos},
  \bibinfo{person}{Mikhail Pavlov}, \bibinfo{person}{Andrew Peng},
  \bibinfo{person}{Adam Perelman}, \bibinfo{person}{Filipe de Avila
  Belbute~Peres}, \bibinfo{person}{Michael Petrov},
  \bibinfo{person}{Henrique~Pond{\'e} de Oliveira~Pinto},
  \bibinfo{person}{Michael Pokorny}, \bibinfo{person}{Michelle Pokrass},
  \bibinfo{person}{Vitchyr~H. Pong}, \bibinfo{person}{Tolly Powell},
  \bibinfo{person}{Alethea Power}, \bibinfo{person}{Boris Power},
  \bibinfo{person}{Elizabeth Proehl}, \bibinfo{person}{Raul Puri},
  \bibinfo{person}{Alec Radford}, \bibinfo{person}{Jack~W. Rae},
  \bibinfo{person}{Aditya Ramesh}, \bibinfo{person}{Cameron Raymond},
  \bibinfo{person}{Francis Real}, \bibinfo{person}{Kendra Rimbach},
  \bibinfo{person}{Carl Ross}, \bibinfo{person}{Bob Rotsted},
  \bibinfo{person}{Henri Roussez}, \bibinfo{person}{Nick Ryder},
  \bibinfo{person}{Mario~D. Saltarelli}, \bibinfo{person}{Ted Sanders},
  \bibinfo{person}{Shibani Santurkar}, \bibinfo{person}{Girish Sastry},
  \bibinfo{person}{Heather Schmidt}, \bibinfo{person}{David Schnurr},
  \bibinfo{person}{John Schulman}, \bibinfo{person}{Daniel Selsam},
  \bibinfo{person}{Kyla Sheppard}, \bibinfo{person}{Toki Sherbakov},
  \bibinfo{person}{Jessica Shieh}, \bibinfo{person}{Sarah Shoker},
  \bibinfo{person}{Pranav Shyam}, \bibinfo{person}{Szymon Sidor},
  \bibinfo{person}{Eric Sigler}, \bibinfo{person}{Maddie Simens},
  \bibinfo{person}{Jordan Sitkin}, \bibinfo{person}{Katarina Slama},
  \bibinfo{person}{Ian Sohl}, \bibinfo{person}{Benjamin Sokolowsky},
  \bibinfo{person}{Yang Song}, \bibinfo{person}{Natalie Staudacher},
  \bibinfo{person}{Felipe~Petroski Such}, \bibinfo{person}{Natalie Summers},
  \bibinfo{person}{Ilya Sutskever}, \bibinfo{person}{Jie Tang},
  \bibinfo{person}{Nikolas~A. Tezak}, \bibinfo{person}{Madeleine Thompson},
  \bibinfo{person}{Phil Tillet}, \bibinfo{person}{Amin Tootoonchian},
  \bibinfo{person}{Elizabeth Tseng}, \bibinfo{person}{Preston Tuggle},
  \bibinfo{person}{Nick Turley}, \bibinfo{person}{Jerry Tworek},
  \bibinfo{person}{Juan Felipe~Cer'on Uribe}, \bibinfo{person}{Andrea Vallone},
  \bibinfo{person}{Arun Vijayvergiya}, \bibinfo{person}{Chelsea Voss},
  \bibinfo{person}{Carroll~L. Wainwright}, \bibinfo{person}{Justin~Jay Wang},
  \bibinfo{person}{Alvin Wang}, \bibinfo{person}{Ben Wang},
  \bibinfo{person}{Jonathan Ward}, \bibinfo{person}{Jason Wei},
  \bibinfo{person}{CJ Weinmann}, \bibinfo{person}{Akila Welihinda},
  \bibinfo{person}{Peter Welinder}, \bibinfo{person}{Jiayi Weng},
  \bibinfo{person}{Lilian Weng}, \bibinfo{person}{Matt Wiethoff},
  \bibinfo{person}{Dave Willner}, \bibinfo{person}{Clemens Winter},
  \bibinfo{person}{Samuel Wolrich}, \bibinfo{person}{Hannah Wong},
  \bibinfo{person}{Lauren Workman}, \bibinfo{person}{Sherwin Wu},
  \bibinfo{person}{Jeff Wu}, \bibinfo{person}{Michael Wu}, \bibinfo{person}{Kai
  Xiao}, \bibinfo{person}{Tao Xu}, \bibinfo{person}{Sarah Yoo},
  \bibinfo{person}{Kevin Yu}, \bibinfo{person}{Qim ing Yuan},
  \bibinfo{person}{Wojciech Zaremba}, \bibinfo{person}{Rowan Zellers},
  \bibinfo{person}{Chong Zhang}, \bibinfo{person}{Marvin Zhang},
  \bibinfo{person}{Shengjia Zhao}, \bibinfo{person}{Tianhao Zheng},
  \bibinfo{person}{Juntang Zhuang}, \bibinfo{person}{William Zhuk}, {and}
  \bibinfo{person}{Barret Zoph}.} \bibinfo{year}{2023}\natexlab{}.
\newblock \showarticletitle{GPT-4 Technical Report}.
\newblock
\urldef\tempurl%
\url{https://api.semanticscholar.org/CorpusID:257532815}
\showURL{%
\tempurl}


\bibitem[Ahmed et~al\mbox{.}(2016)]%
        {ahmed_can_2016}
\bibfield{author}{\bibinfo{person}{Iftekhar Ahmed}, \bibinfo{person}{Rahul
  Gopinath}, \bibinfo{person}{Caius Brindescu}, \bibinfo{person}{Alex Groce},
  {and} \bibinfo{person}{Carlos Jensen}.} \bibinfo{year}{2016}\natexlab{}.
\newblock \showarticletitle{Can testedness be effectively measured?}. In
  \bibinfo{booktitle}{\emph{Proceedings of the 2016 24th {ACM} {SIGSOFT}
  {International} {Symposium} on {Foundations} of {Software} {Engineering}}}.
  \bibinfo{publisher}{ACM}, \bibinfo{address}{Seattle WA USA},
  \bibinfo{pages}{547--558}.
\newblock
\showISBNx{978-1-4503-4218-6}
\href{https://doi.org/10.1145/2950290.2950324}{doi:\nolinkurl{10.1145/2950290.2950324}}


\bibitem[Alagarsamy et~al\mbox{.}(2024)]%
        {alagarsamy_a3test_2023}
\bibfield{author}{\bibinfo{person}{Saranya Alagarsamy},
  \bibinfo{person}{Chakkrit Tantithamthavorn}, {and} \bibinfo{person}{Aldeida
  Aleti}.} \bibinfo{year}{2024}\natexlab{}.
\newblock \showarticletitle{A3Test: Assertion-Augmented Automated Test case
  generation}.
\newblock \bibinfo{journal}{\emph{Inf. Softw. Technol.}} \bibinfo{volume}{176},
  \bibinfo{number}{C} (\bibinfo{date}{Dec.} \bibinfo{year}{2024}),
  \bibinfo{numpages}{15}~pages.
\newblock
\showISSN{0950-5849}
\href{https://doi.org/10.1016/j.infsof.2024.107565}{doi:\nolinkurl{10.1016/j.infsof.2024.107565}}


\bibitem[Alshahwan et~al\mbox{.}(2024a)]%
        {alshahwan_automated_2024}
\bibfield{author}{\bibinfo{person}{Nadia Alshahwan}, \bibinfo{person}{Jubin
  Chheda}, \bibinfo{person}{Anastasia Finogenova}, \bibinfo{person}{Beliz
  Gokkaya}, \bibinfo{person}{Mark Harman}, \bibinfo{person}{Inna Harper},
  \bibinfo{person}{Alexandru Marginean}, \bibinfo{person}{Shubho Sengupta},
  {and} \bibinfo{person}{Eddy Wang}.} \bibinfo{year}{2024}\natexlab{a}.
\newblock \showarticletitle{Automated Unit Test Improvement using Large
  Language Models at Meta}. In \bibinfo{booktitle}{\emph{Companion Proceedings
  of the 32nd ACM International Conference on the Foundations of Software
  Engineering}} (Porto de Galinhas, Brazil) \emph{(\bibinfo{series}{FSE
  2024})}. \bibinfo{publisher}{Association for Computing Machinery},
  \bibinfo{address}{New York, NY, USA}, \bibinfo{pages}{185–196}.
\newblock
\showISBNx{9798400706585}
\href{https://doi.org/10.1145/3663529.3663839}{doi:\nolinkurl{10.1145/3663529.3663839}}


\bibitem[Alshahwan et~al\mbox{.}(2024b)]%
        {alshahwan_assured_2024}
\bibfield{author}{\bibinfo{person}{Nadia Alshahwan}, \bibinfo{person}{Mark
  Harman}, \bibinfo{person}{Inna Harper}, \bibinfo{person}{Alexandru
  Marginean}, \bibinfo{person}{Shubho Sengupta}, {and} \bibinfo{person}{Eddy
  Wang}.} \bibinfo{year}{2024}\natexlab{b}.
\newblock \bibinfo{title}{Assured {LLM}-{Based} {Software} {Engineering}}.
\newblock
\urldef\tempurl%
\url{http://arxiv.org/abs/2402.04380}
\showURL{%
\tempurl}
\newblock
\shownote{arXiv:2402.04380 [cs]}.


\bibitem[Bavota et~al\mbox{.}(2012)]%
        {bavota_empirical_2012}
\bibfield{author}{\bibinfo{person}{Gabriele Bavota}, \bibinfo{person}{Abdallah
  Qusef}, \bibinfo{person}{Rocco Oliveto}, \bibinfo{person}{Andrea De~Lucia},
  {and} \bibinfo{person}{David Binkley}.} \bibinfo{year}{2012}\natexlab{}.
\newblock \showarticletitle{An empirical analysis of the distribution of unit
  test smells and their impact on software maintenance}. In
  \bibinfo{booktitle}{\emph{2012 28th {IEEE} {International} {Conference} on
  {Software} {Maintenance} ({ICSM})}}. \bibinfo{pages}{56--65}.
\newblock
\href{https://doi.org/10.1109/ICSM.2012.6405253}{doi:\nolinkurl{10.1109/ICSM.2012.6405253}}
\newblock
\shownote{ISSN: 1063-6773}.


\bibitem[Brown et~al\mbox{.}(2020)]%
        {brown_language_2020}
\bibfield{author}{\bibinfo{person}{Tom~B. Brown}, \bibinfo{person}{Benjamin
  Mann}, \bibinfo{person}{Nick Ryder}, \bibinfo{person}{Melanie Subbiah},
  \bibinfo{person}{Jared Kaplan}, \bibinfo{person}{Prafulla Dhariwal},
  \bibinfo{person}{Arvind Neelakantan}, \bibinfo{person}{Pranav Shyam},
  \bibinfo{person}{Girish Sastry}, \bibinfo{person}{Amanda Askell},
  \bibinfo{person}{Sandhini Agarwal}, \bibinfo{person}{Ariel Herbert-Voss},
  \bibinfo{person}{Gretchen Krueger}, \bibinfo{person}{Tom Henighan},
  \bibinfo{person}{Rewon Child}, \bibinfo{person}{Aditya Ramesh},
  \bibinfo{person}{Daniel~M. Ziegler}, \bibinfo{person}{Jeffrey Wu},
  \bibinfo{person}{Clemens Winter}, \bibinfo{person}{Christopher Hesse},
  \bibinfo{person}{Mark Chen}, \bibinfo{person}{Eric Sigler},
  \bibinfo{person}{Mateusz Litwin}, \bibinfo{person}{Scott Gray},
  \bibinfo{person}{Benjamin Chess}, \bibinfo{person}{Jack Clark},
  \bibinfo{person}{Christopher Berner}, \bibinfo{person}{Sam McCandlish},
  \bibinfo{person}{Alec Radford}, \bibinfo{person}{Ilya Sutskever}, {and}
  \bibinfo{person}{Dario Amodei}.} \bibinfo{year}{2020}\natexlab{}.
\newblock \showarticletitle{Language models are few-shot learners}. In
  \bibinfo{booktitle}{\emph{Proceedings of the 34th International Conference on
  Neural Information Processing Systems}} (Vancouver, BC, Canada)
  \emph{(\bibinfo{series}{NIPS '20})}. \bibinfo{publisher}{Curran Associates
  Inc.}, \bibinfo{address}{Red Hook, NY, USA}, Article
  \bibinfo{articleno}{159}, \bibinfo{numpages}{25}~pages.
\newblock
\showISBNx{9781713829546}


\bibitem[Chen et~al\mbox{.}(2021)]%
        {chen_evaluating_2021}
\bibfield{author}{\bibinfo{person}{Mark Chen}, \bibinfo{person}{Jerry Tworek},
  \bibinfo{person}{Heewoo Jun}, \bibinfo{person}{Qiming Yuan},
  \bibinfo{person}{Henrique Ponde de~Oliveira Pinto}, \bibinfo{person}{Jared
  Kaplan}, \bibinfo{person}{Harri Edwards}, \bibinfo{person}{Yuri Burda},
  \bibinfo{person}{Nicholas Joseph}, \bibinfo{person}{Greg Brockman},
  \bibinfo{person}{Alex Ray}, \bibinfo{person}{Raul Puri},
  \bibinfo{person}{Gretchen Krueger}, \bibinfo{person}{Michael Petrov},
  \bibinfo{person}{Heidy Khlaaf}, \bibinfo{person}{Girish Sastry},
  \bibinfo{person}{Pamela Mishkin}, \bibinfo{person}{Brooke Chan},
  \bibinfo{person}{Scott Gray}, \bibinfo{person}{Nick Ryder},
  \bibinfo{person}{Mikhail Pavlov}, \bibinfo{person}{Alethea Power},
  \bibinfo{person}{Lukasz Kaiser}, \bibinfo{person}{Mohammad Bavarian},
  \bibinfo{person}{Clemens Winter}, \bibinfo{person}{Philippe Tillet},
  \bibinfo{person}{Felipe~Petroski Such}, \bibinfo{person}{Dave Cummings},
  \bibinfo{person}{Matthias Plappert}, \bibinfo{person}{Fotios Chantzis},
  \bibinfo{person}{Elizabeth Barnes}, \bibinfo{person}{Ariel Herbert-Voss},
  \bibinfo{person}{William~Hebgen Guss}, \bibinfo{person}{Alex Nichol},
  \bibinfo{person}{Alex Paino}, \bibinfo{person}{Nikolas Tezak},
  \bibinfo{person}{Jie Tang}, \bibinfo{person}{Igor Babuschkin},
  \bibinfo{person}{Suchir Balaji}, \bibinfo{person}{Shantanu Jain},
  \bibinfo{person}{William Saunders}, \bibinfo{person}{Christopher Hesse},
  \bibinfo{person}{Andrew~N. Carr}, \bibinfo{person}{Jan Leike},
  \bibinfo{person}{Josh Achiam}, \bibinfo{person}{Vedant Misra},
  \bibinfo{person}{Evan Morikawa}, \bibinfo{person}{Alec Radford},
  \bibinfo{person}{Matthew Knight}, \bibinfo{person}{Miles Brundage},
  \bibinfo{person}{Mira Murati}, \bibinfo{person}{Katie Mayer},
  \bibinfo{person}{Peter Welinder}, \bibinfo{person}{Bob McGrew},
  \bibinfo{person}{Dario Amodei}, \bibinfo{person}{Sam McCandlish},
  \bibinfo{person}{Ilya Sutskever}, {and} \bibinfo{person}{Wojciech Zaremba}.}
  \bibinfo{year}{2021}\natexlab{}.
\newblock \bibinfo{title}{Evaluating Large Language Models Trained on Code}.
\newblock
\showeprint[arxiv]{2107.03374}~[cs.LG]


\bibitem[Chudic and {u}l~\c Cal{\i}kl{\i}(2025)]%
        {anonymous_2025_15561007}
\bibfield{author}{\bibinfo{person}{Alex Chudic} {and} \bibinfo{person}{G\"
  {u}l~\c Cal{\i}kl{\i}}.} \bibinfo{year}{2025}\natexlab{}.
\newblock \bibinfo{title}{Automated Test Suite Enhancement Using Large Language
  Models With Few-shot Prompting}.
\newblock
\href{https://doi.org/10.5281/zenodo.15561007}{doi:\nolinkurl{10.5281/zenodo.15561007}}


\bibitem[Du et~al\mbox{.}(2023)]%
        {du_classeval_2023}
\bibfield{author}{\bibinfo{person}{Xueying Du}, \bibinfo{person}{Mingwei Liu},
  \bibinfo{person}{Kaixin Wang}, \bibinfo{person}{Hanlin Wang},
  \bibinfo{person}{Junwei Liu}, \bibinfo{person}{Yixuan Chen},
  \bibinfo{person}{Jiayi Feng}, \bibinfo{person}{Chaofeng Sha},
  \bibinfo{person}{Xin Peng}, {and} \bibinfo{person}{Yiling Lou}.}
  \bibinfo{year}{2023}\natexlab{}.
\newblock \bibinfo{title}{ClassEval: A Manually-Crafted Benchmark for
  Evaluating LLMs on Class-level Code Generation}.
\newblock
\showeprint[arxiv]{2308.01861}~[cs.CL]


\bibitem[Döderlein et~al\mbox{.}(2023)]%
        {doderlein_piloting_2023}
\bibfield{author}{\bibinfo{person}{Jean-Baptiste Döderlein},
  \bibinfo{person}{Mathieu Acher}, \bibinfo{person}{Djamel~Eddine Khelladi},
  {and} \bibinfo{person}{Benoit Combemale}.} \bibinfo{year}{2023}\natexlab{}.
\newblock \bibinfo{title}{Piloting {Copilot} and {Codex}: {Hot} {Temperature},
  {Cold} {Prompts}, or {Black} {Magic}?}
\newblock
\urldef\tempurl%
\url{https://dx.doi.org/10.2139/ssrn.4496380}
\showURL{%
\tempurl}


\bibitem[Fan et~al\mbox{.}(2023)]%
        {fan_large_2023}
\bibfield{author}{\bibinfo{person}{Angela Fan}, \bibinfo{person}{Beliz
  Gokkaya}, \bibinfo{person}{Mark Harman}, \bibinfo{person}{Mitya Lyubarskiy},
  \bibinfo{person}{Shubho Sengupta}, \bibinfo{person}{Shin Yoo}, {and}
  \bibinfo{person}{Jie~M. Zhang}.} \bibinfo{year}{2023}\natexlab{}.
\newblock \showarticletitle{Large Language Models for Software Engineering:
  Survey and Open Problems}. In \bibinfo{booktitle}{\emph{2023 IEEE/ACM
  International Conference on Software Engineering: Future of Software
  Engineering (ICSE-FoSE)}}. \bibinfo{pages}{31--53}.
\newblock
\href{https://doi.org/10.1109/ICSE-FoSE59343.2023.00008}{doi:\nolinkurl{10.1109/ICSE-FoSE59343.2023.00008}}


\bibitem[Fraser and Arcuri(2011)]%
        {fraser_evosuite_2011}
\bibfield{author}{\bibinfo{person}{Gordon Fraser} {and} \bibinfo{person}{Andrea
  Arcuri}.} \bibinfo{year}{2011}\natexlab{}.
\newblock \showarticletitle{{EvoSuite}: automatic test suite generation for
  object-oriented software}. In \bibinfo{booktitle}{\emph{Proceedings of the
  19th {ACM} {SIGSOFT} symposium and the 13th {European} conference on
  {Foundations} of software engineering}}. \bibinfo{publisher}{ACM},
  \bibinfo{address}{Szeged Hungary}, \bibinfo{pages}{416--419}.
\newblock
\showISBNx{978-1-4503-0443-6}
\href{https://doi.org/10.1145/2025113.2025179}{doi:\nolinkurl{10.1145/2025113.2025179}}


\bibitem[Harman et~al\mbox{.}(2012)]%
        {harman_search-based_2012}
\bibfield{author}{\bibinfo{person}{Mark Harman}, \bibinfo{person}{S.~Afshin
  Mansouri}, {and} \bibinfo{person}{Yuanyuan Zhang}.}
  \bibinfo{year}{2012}\natexlab{}.
\newblock \showarticletitle{Search-based software engineering: {Trends},
  techniques and applications}.
\newblock \bibinfo{journal}{\emph{ACM Comput. Surv.}} \bibinfo{volume}{45},
  \bibinfo{number}{1} (\bibinfo{date}{Dec.} \bibinfo{year}{2012}),
  \bibinfo{pages}{11:1--11:61}.
\newblock
\showISSN{0360-0300}
\href{https://doi.org/10.1145/2379776.2379787}{doi:\nolinkurl{10.1145/2379776.2379787}}


\bibitem[Harman and McMinn(2010)]%
        {harman_theoretical_2010}
\bibfield{author}{\bibinfo{person}{Mark Harman} {and} \bibinfo{person}{Phil
  McMinn}.} \bibinfo{year}{2010}\natexlab{}.
\newblock \showarticletitle{A {Theoretical} and {Empirical} {Study} of
  {Search}-{Based} {Testing}: {Local}, {Global}, and {Hybrid} {Search}}.
\newblock \bibinfo{journal}{\emph{IEEE Transactions on Software Engineering}}
  \bibinfo{volume}{36}, \bibinfo{number}{2} (\bibinfo{date}{March}
  \bibinfo{year}{2010}), \bibinfo{pages}{226--247}.
\newblock
\showISSN{1939-3520}
\href{https://doi.org/10.1109/TSE.2009.71}{doi:\nolinkurl{10.1109/TSE.2009.71}}
\newblock
\shownote{Conference Name: IEEE Transactions on Software Engineering}.


\bibitem[Inozemtseva and Holmes(2014)]%
        {inozemtseva_coverage_2014}
\bibfield{author}{\bibinfo{person}{Laura Inozemtseva} {and}
  \bibinfo{person}{Reid Holmes}.} \bibinfo{year}{2014}\natexlab{}.
\newblock \showarticletitle{Coverage is not strongly correlated with test suite
  effectiveness}. In \bibinfo{booktitle}{\emph{Proceedings of the 36th
  {International} {Conference} on {Software} {Engineering}}}.
  \bibinfo{publisher}{ACM}, \bibinfo{address}{Hyderabad India},
  \bibinfo{pages}{435--445}.
\newblock
\showISBNx{978-1-4503-2756-5}
\href{https://doi.org/10.1145/2568225.2568271}{doi:\nolinkurl{10.1145/2568225.2568271}}


\bibitem[Kojima et~al\mbox{.}(2022)]%
        {kojima_large_2023}
\bibfield{author}{\bibinfo{person}{Takeshi Kojima},
  \bibinfo{person}{Shixiang~Shane Gu}, \bibinfo{person}{Machel Reid},
  \bibinfo{person}{Yutaka Matsuo}, {and} \bibinfo{person}{Yusuke Iwasawa}.}
  \bibinfo{year}{2022}\natexlab{}.
\newblock \showarticletitle{Large language models are zero-shot reasoners}. In
  \bibinfo{booktitle}{\emph{Proceedings of the 36th International Conference on
  Neural Information Processing Systems}} (New Orleans, LA, USA)
  \emph{(\bibinfo{series}{NIPS '22})}. \bibinfo{publisher}{Curran Associates
  Inc.}, \bibinfo{address}{Red Hook, NY, USA}, Article
  \bibinfo{articleno}{1613}, \bibinfo{numpages}{15}~pages.
\newblock
\showISBNx{9781713871088}


\bibitem[Lemieux et~al\mbox{.}(2023)]%
        {lemieux_codamosa_2023}
\bibfield{author}{\bibinfo{person}{Caroline Lemieux},
  \bibinfo{person}{Jeevana~Priya Inala}, \bibinfo{person}{Shuvendu~K. Lahiri},
  {and} \bibinfo{person}{Siddhartha Sen}.} \bibinfo{year}{2023}\natexlab{}.
\newblock \showarticletitle{{CodaMosa}: {Escaping} {Coverage} {Plateaus} in
  {Test} {Generation} with {Pre}-trained {Large} {Language} {Models}}. In
  \bibinfo{booktitle}{\emph{2023 {IEEE}/{ACM} 45th {International} {Conference}
  on {Software} {Engineering} ({ICSE})}}. \bibinfo{pages}{919--931}.
\newblock
\href{https://doi.org/10.1109/ICSE48619.2023.00085}{doi:\nolinkurl{10.1109/ICSE48619.2023.00085}}
\newblock
\shownote{ISSN: 1558-1225}.


\bibitem[Liu et~al\mbox{.}(2023)]%
        {liu_is_2023}
\bibfield{author}{\bibinfo{person}{Jiawei Liu}, \bibinfo{person}{Chunqiu~Steven
  Xia}, \bibinfo{person}{Yuyao Wang}, {and} \bibinfo{person}{Lingming Zhang}.}
  \bibinfo{year}{2023}\natexlab{}.
\newblock \showarticletitle{Is your code generated by ChatGPT really correct?
  rigorous evaluation of large language models for code generation}. In
  \bibinfo{booktitle}{\emph{Proceedings of the 37th International Conference on
  Neural Information Processing Systems}} (New Orleans, LA, USA)
  \emph{(\bibinfo{series}{NIPS '23})}. \bibinfo{publisher}{Curran Associates
  Inc.}, \bibinfo{address}{Red Hook, NY, USA}, Article
  \bibinfo{articleno}{943}, \bibinfo{numpages}{15}~pages.
\newblock


\bibitem[Nashid et~al\mbox{.}(2023)]%
        {noor_retrieval_2023}
\bibfield{author}{\bibinfo{person}{Noor Nashid}, \bibinfo{person}{Mifta
  Sintaha}, {and} \bibinfo{person}{Ali Mesbah}.}
  \bibinfo{year}{2023}\natexlab{}.
\newblock \showarticletitle{Retrieval-Based Prompt Selection for Code-Related
  Few-Shot Learning}. In \bibinfo{booktitle}{\emph{Proceedings of the 45th
  International Conference on Software Engineering}} (Melbourne, Victoria,
  Australia) \emph{(\bibinfo{series}{ICSE '23})}. \bibinfo{publisher}{IEEE
  Press}, \bibinfo{pages}{2450–2462}.
\newblock
\showISBNx{9781665457019}
\href{https://doi.org/10.1109/ICSE48619.2023.00205}{doi:\nolinkurl{10.1109/ICSE48619.2023.00205}}


\bibitem[Nie et~al\mbox{.}(2023)]%
        {nie_learning_2023}
\bibfield{author}{\bibinfo{person}{Pengyu Nie}, \bibinfo{person}{Rahul
  Banerjee}, \bibinfo{person}{Junyi~Jessy Li}, \bibinfo{person}{Raymond~J.
  Mooney}, {and} \bibinfo{person}{Milos Gligoric}.}
  \bibinfo{year}{2023}\natexlab{}.
\newblock \showarticletitle{Learning Deep Semantics for Test Completion}. In
  \bibinfo{booktitle}{\emph{Proceedings of the 45th International Conference on
  Software Engineering}} (Melbourne, Victoria, Australia)
  \emph{(\bibinfo{series}{ICSE '23})}. \bibinfo{publisher}{IEEE Press},
  \bibinfo{pages}{2111–2123}.
\newblock
\showISBNx{9781665457019}
\href{https://doi.org/10.1109/ICSE48619.2023.00178}{doi:\nolinkurl{10.1109/ICSE48619.2023.00178}}


\bibitem[Ouyang et~al\mbox{.}(2025)]%
        {ouyang_llm_2023}
\bibfield{author}{\bibinfo{person}{Shuyin Ouyang}, \bibinfo{person}{Jie~M.
  Zhang}, \bibinfo{person}{Mark Harman}, {and} \bibinfo{person}{Meng Wang}.}
  \bibinfo{year}{2025}\natexlab{}.
\newblock \showarticletitle{An Empirical Study of the Non-Determinism of
  ChatGPT in Code Generation}.
\newblock \bibinfo{journal}{\emph{ACM Trans. Softw. Eng. Methodol.}}
  \bibinfo{volume}{34}, \bibinfo{number}{2}, Article \bibinfo{articleno}{42}
  (\bibinfo{date}{Jan.} \bibinfo{year}{2025}), \bibinfo{numpages}{28}~pages.
\newblock
\showISSN{1049-331X}
\href{https://doi.org/10.1145/3697010}{doi:\nolinkurl{10.1145/3697010}}


\bibitem[Ouédraogo et~al\mbox{.}(2024)]%
        {ouedraogo_large-scale_2024}
\bibfield{author}{\bibinfo{person}{Wendkûuni Ouédraogo},
  \bibinfo{person}{Kader Kaboré}, \bibinfo{person}{Haoye Tian},
  \bibinfo{person}{Yewei Song}, \bibinfo{person}{Anil Koyuncu},
  \bibinfo{person}{Jacques Klein}, \bibinfo{person}{David Lo}, {and}
  \bibinfo{person}{Tegawendé Bissyandé}.} \bibinfo{year}{2024}\natexlab{}.
\newblock \bibinfo{title}{Large-scale, Independent and Comprehensive study of
  the power of LLMs for test case generation}.
\newblock
\href{https://doi.org/10.48550/arXiv.2407.00225}{doi:\nolinkurl{10.48550/arXiv.2407.00225}}


\bibitem[Pacheco and Ernst(2007)]%
        {pacheco_randoop_2007}
\bibfield{author}{\bibinfo{person}{Carlos Pacheco} {and}
  \bibinfo{person}{Michael~D. Ernst}.} \bibinfo{year}{2007}\natexlab{}.
\newblock \showarticletitle{Randoop: feedback-directed random testing for
  {Java}}. In \bibinfo{booktitle}{\emph{Companion to the 22nd {ACM} {SIGPLAN}
  conference on {Object}-oriented programming systems and applications
  companion}} \emph{(\bibinfo{series}{{OOPSLA} '07})}.
  \bibinfo{publisher}{Association for Computing Machinery},
  \bibinfo{address}{New York, NY, USA}, \bibinfo{pages}{815--816}.
\newblock
\showISBNx{978-1-59593-865-7}
\href{https://doi.org/10.1145/1297846.1297902}{doi:\nolinkurl{10.1145/1297846.1297902}}


\bibitem[Pacheco et~al\mbox{.}(2007)]%
        {pacheco_feedback-directed_2007}
\bibfield{author}{\bibinfo{person}{Carlos Pacheco},
  \bibinfo{person}{Shuvendu~K. Lahiri}, \bibinfo{person}{Michael~D. Ernst},
  {and} \bibinfo{person}{Thomas Ball}.} \bibinfo{year}{2007}\natexlab{}.
\newblock \showarticletitle{Feedback-{Directed} {Random} {Test} {Generation}}.
  In \bibinfo{booktitle}{\emph{29th {International} {Conference} on {Software}
  {Engineering} ({ICSE}'07)}}. \bibinfo{pages}{75--84}.
\newblock
\href{https://doi.org/10.1109/ICSE.2007.37}{doi:\nolinkurl{10.1109/ICSE.2007.37}}
\newblock
\shownote{ISSN: 1558-1225}.


\bibitem[Panichella et~al\mbox{.}(2020)]%
        {panichella_testsmells_2020}
\bibfield{author}{\bibinfo{person}{Annibale Panichella},
  \bibinfo{person}{Sebastiano Panichella}, \bibinfo{person}{Gordon Fraser},
  \bibinfo{person}{Anand~Ashok Sawant}, {and} \bibinfo{person}{Vincent~J.
  Hellendoorn}.} \bibinfo{year}{2020}\natexlab{}.
\newblock \showarticletitle{Revisiting Test Smells in Automatically Generated
  Tests: Limitations, Pitfalls, and Opportunities}. In
  \bibinfo{booktitle}{\emph{2020 IEEE International Conference on Software
  Maintenance and Evolution (ICSME)}}. \bibinfo{pages}{523--533}.
\newblock
\href{https://doi.org/10.1109/ICSME46990.2020.00056}{doi:\nolinkurl{10.1109/ICSME46990.2020.00056}}


\bibitem[Roy et~al\mbox{.}(2021)]%
        {roy_deepTC_2021}
\bibfield{author}{\bibinfo{person}{Devjeet Roy}, \bibinfo{person}{Ziyi Zhang},
  \bibinfo{person}{Maggie Ma}, \bibinfo{person}{Venera Arnaoudova},
  \bibinfo{person}{Annibale Panichella}, \bibinfo{person}{Sebastiano
  Panichella}, \bibinfo{person}{Danielle Gonzalez}, {and}
  \bibinfo{person}{Mehdi Mirakhorli}.} \bibinfo{year}{2021}\natexlab{}.
\newblock \showarticletitle{DeepTC-enhancer: improving the readability of
  automatically generated tests}. In \bibinfo{booktitle}{\emph{Proceedings of
  the 35th IEEE/ACM International Conference on Automated Software
  Engineering}} (Virtual Event, Australia) \emph{(\bibinfo{series}{ASE '20})}.
  \bibinfo{publisher}{Association for Computing Machinery},
  \bibinfo{address}{New York, NY, USA}, \bibinfo{pages}{287–298}.
\newblock
\showISBNx{9781450367684}
\href{https://doi.org/10.1145/3324884.3416622}{doi:\nolinkurl{10.1145/3324884.3416622}}


\bibitem[Schäfer et~al\mbox{.}(2024)]%
        {schafer_adaptive_2023}
\bibfield{author}{\bibinfo{person}{Max Schäfer}, \bibinfo{person}{Sarah Nadi},
  \bibinfo{person}{Aryaz Eghbali}, {and} \bibinfo{person}{Frank Tip}.}
  \bibinfo{year}{2024}\natexlab{}.
\newblock \showarticletitle{An Empirical Evaluation of Using Large Language
  Models for Automated Unit Test Generation}.
\newblock \bibinfo{journal}{\emph{IEEE Transactions on Software Engineering}}
  \bibinfo{volume}{50}, \bibinfo{number}{1} (\bibinfo{year}{2024}),
  \bibinfo{pages}{85--105}.
\newblock
\href{https://doi.org/10.1109/TSE.2023.3334955}{doi:\nolinkurl{10.1109/TSE.2023.3334955}}


\bibitem[Siddiq et~al\mbox{.}(2024a)]%
        {siddiq_using_2024}
\bibfield{author}{\bibinfo{person}{Mohammed~Latif Siddiq},
  \bibinfo{person}{Joanna~Cecilia Da~Silva~Santos},
  \bibinfo{person}{Ridwanul~Hasan Tanvir}, \bibinfo{person}{Noshin Ulfat},
  \bibinfo{person}{Fahmid Al~Rifat}, {and} \bibinfo{person}{Vin\'{\i}cius
  Carvalho~Lopes}.} \bibinfo{year}{2024}\natexlab{a}.
\newblock \showarticletitle{Using Large Language Models to Generate JUnit
  Tests: An Empirical Study}. In \bibinfo{booktitle}{\emph{Proceedings of the
  28th International Conference on Evaluation and Assessment in Software
  Engineering}} (Salerno, Italy) \emph{(\bibinfo{series}{EASE '24})}.
  \bibinfo{publisher}{Association for Computing Machinery},
  \bibinfo{address}{New York, NY, USA}, \bibinfo{pages}{313–322}.
\newblock
\showISBNx{9798400717017}
\href{https://doi.org/10.1145/3661167.3661216}{doi:\nolinkurl{10.1145/3661167.3661216}}


\bibitem[Siddiq et~al\mbox{.}(2024b)]%
        {siddiq_fault_2024}
\bibfield{author}{\bibinfo{person}{Mohammed~Latif Siddiq},
  \bibinfo{person}{Simantika Dristi}, \bibinfo{person}{Joy Saha}, {and}
  \bibinfo{person}{Joanna C.~S. Santos}.} \bibinfo{year}{2024}\natexlab{b}.
\newblock \bibinfo{title}{The {Fault} in our {Stars}: {Quality} {Assessment} of
  {Code} {Generation} {Benchmarks}}.
\newblock
\href{https://doi.org/10.48550/arXiv.2404.10155}{doi:\nolinkurl{10.48550/arXiv.2404.10155}}
\newblock
\shownote{arXiv:2404.10155}.


\bibitem[{SonarSource SA}(2025a)]%
        {sonarqube-cloud}
\bibfield{author}{\bibinfo{person}{{SonarSource SA}}.}
  \bibinfo{year}{2025}\natexlab{a}.
\newblock \bibinfo{title}{SonarQube Cloud}.
\newblock \bibinfo{howpublished}{\url{https://sonarcloud.io}}.
\newblock


\bibitem[{SonarSource SA}(2025b)]%
        {sonarscanner-cli}
\bibfield{author}{\bibinfo{person}{{SonarSource SA}}.}
  \bibinfo{year}{2025}\natexlab{b}.
\newblock \bibinfo{title}{SonarScanner CLI}.
\newblock
  \bibinfo{howpublished}{\url{https://github.com/SonarSource/sonar-scanner-cli}}.
\newblock
\newblock
\shownote{Version 5.x}.


\bibitem[{SonarSource SA}(2025c)]%
        {sqale-rating}
\bibfield{author}{\bibinfo{person}{{SonarSource SA}}.}
  \bibinfo{year}{2025}\natexlab{c}.
\newblock \bibinfo{title}{Squale rating of SonarQube}.
\newblock
  \bibinfo{howpublished}{\url{https://docs.sonarsource.com/sonarqube-server/9.9/user-guide/metric-definitions}}.
\newblock


\bibitem[Tufano et~al\mbox{.}(2020)]%
        {tufano_unit_2021}
\bibfield{author}{\bibinfo{person}{Michele Tufano}, \bibinfo{person}{Dawn
  Drain}, \bibinfo{person}{Alexey Svyatkovskiy}, \bibinfo{person}{Shao Deng},
  {and} \bibinfo{person}{Neel Sundaresan}.} \bibinfo{year}{2020}\natexlab{}.
\newblock \bibinfo{title}{Unit Test Case Generation with Transformers}.
\newblock
\href{https://doi.org/10.48550/arXiv.2009.05617}{doi:\nolinkurl{10.48550/arXiv.2009.05617}}


\bibitem[Tufano et~al\mbox{.}(2015)]%
        {tufano_when_2015}
\bibfield{author}{\bibinfo{person}{Michele Tufano}, \bibinfo{person}{Fabio
  Palomba}, \bibinfo{person}{Gabriele Bavota}, \bibinfo{person}{Rocco Oliveto},
  \bibinfo{person}{Massimiliano Di~Penta}, \bibinfo{person}{Andrea De~Lucia},
  {and} \bibinfo{person}{Denys Poshyvanyk}.} \bibinfo{year}{2015}\natexlab{}.
\newblock \showarticletitle{When and {Why} {Your} {Code} {Starts} to {Smell}
  {Bad}}. In \bibinfo{booktitle}{\emph{2015 {IEEE}/{ACM} 37th {IEEE}
  {International} {Conference} on {Software} {Engineering}}}.
  \bibinfo{publisher}{IEEE}, \bibinfo{address}{Florence, Italy},
  \bibinfo{pages}{403--414}.
\newblock
\showISBNx{978-1-4799-1934-5}
\href{https://doi.org/10.1109/ICSE.2015.59}{doi:\nolinkurl{10.1109/ICSE.2015.59}}


\bibitem[Vaswani et~al\mbox{.}(2017)]%
        {vaswani_attention_2023}
\bibfield{author}{\bibinfo{person}{Ashish Vaswani}, \bibinfo{person}{Noam
  Shazeer}, \bibinfo{person}{Niki Parmar}, \bibinfo{person}{Jakob Uszkoreit},
  \bibinfo{person}{Llion Jones}, \bibinfo{person}{Aidan~N. Gomez},
  \bibinfo{person}{\L{}ukasz Kaiser}, {and} \bibinfo{person}{Illia
  Polosukhin}.} \bibinfo{year}{2017}\natexlab{}.
\newblock \showarticletitle{Attention is all you need}. In
  \bibinfo{booktitle}{\emph{Proceedings of the 31st International Conference on
  Neural Information Processing Systems}} (Long Beach, California, USA)
  \emph{(\bibinfo{series}{NIPS'17})}. \bibinfo{publisher}{Curran Associates
  Inc.}, \bibinfo{address}{Red Hook, NY, USA}, \bibinfo{pages}{6000–6010}.
\newblock
\showISBNx{9781510860964}


\bibitem[Wang et~al\mbox{.}(2024)]%
        {wang_software_2023}
\bibfield{author}{\bibinfo{person}{Junjie Wang}, \bibinfo{person}{Yuchao
  Huang}, \bibinfo{person}{Chunyang Chen}, \bibinfo{person}{Zhe Liu},
  \bibinfo{person}{Song Wang}, {and} \bibinfo{person}{Qing Wang}.}
  \bibinfo{year}{2024}\natexlab{}.
\newblock \showarticletitle{Software Testing With Large Language Models:
  Survey, Landscape, and Vision}.
\newblock \bibinfo{journal}{\emph{IEEE Trans. Softw. Eng.}}
  \bibinfo{volume}{50}, \bibinfo{number}{4} (\bibinfo{date}{April}
  \bibinfo{year}{2024}), \bibinfo{pages}{911–936}.
\newblock
\showISSN{0098-5589}
\href{https://doi.org/10.1109/TSE.2024.3368208}{doi:\nolinkurl{10.1109/TSE.2024.3368208}}


\bibitem[Wei et~al\mbox{.}(2022)]%
        {wei_chain--thought_2023}
\bibfield{author}{\bibinfo{person}{Jason Wei}, \bibinfo{person}{Xuezhi Wang},
  \bibinfo{person}{Dale Schuurmans}, \bibinfo{person}{Maarten Bosma},
  \bibinfo{person}{Brian Ichter}, \bibinfo{person}{Fei Xia},
  \bibinfo{person}{Ed~H. Chi}, \bibinfo{person}{Quoc~V. Le}, {and}
  \bibinfo{person}{Denny Zhou}.} \bibinfo{year}{2022}\natexlab{}.
\newblock \showarticletitle{Chain-of-thought prompting elicits reasoning in
  large language models}. In \bibinfo{booktitle}{\emph{Proceedings of the 36th
  International Conference on Neural Information Processing Systems}} (New
  Orleans, LA, USA) \emph{(\bibinfo{series}{NIPS '22})}.
  \bibinfo{publisher}{Curran Associates Inc.}, \bibinfo{address}{Red Hook, NY,
  USA}, Article \bibinfo{articleno}{1800}, \bibinfo{numpages}{14}~pages.
\newblock
\showISBNx{9781713871088}


\bibitem[White et~al\mbox{.}(2023)]%
        {white_prompt_2023}
\bibfield{author}{\bibinfo{person}{Jules White}, \bibinfo{person}{Quchen Fu},
  \bibinfo{person}{Sam Hays}, \bibinfo{person}{Michael Sandborn},
  \bibinfo{person}{Carlos Olea}, \bibinfo{person}{Henry Gilbert},
  \bibinfo{person}{Ashraf Elnashar}, \bibinfo{person}{Jesse Spencer-Smith},
  {and} \bibinfo{person}{Douglas~C. Schmidt}.} \bibinfo{year}{2023}\natexlab{}.
\newblock \showarticletitle{A Prompt Pattern Catalog to Enhance Prompt
  Engineering with ChatGPT}. In \bibinfo{booktitle}{\emph{Proceedings of the
  30th Conference on Pattern Languages of Programs}} (Monticello, IL, USA)
  \emph{(\bibinfo{series}{PLoP '23})}. \bibinfo{publisher}{The Hillside Group},
  \bibinfo{address}{USA}, Article \bibinfo{articleno}{5},
  \bibinfo{numpages}{31}~pages.
\newblock
\showISBNx{9781941652190}


\bibitem[White et~al\mbox{.}(2024)]%
        {white_chatgpt_2023}
\bibfield{author}{\bibinfo{person}{Jules White}, \bibinfo{person}{Sam Hays},
  \bibinfo{person}{Quchen Fu}, \bibinfo{person}{Jesse Spencer-Smith}, {and}
  \bibinfo{person}{Douglas~C. Schmidt}.} \bibinfo{year}{2024}\natexlab{}.
\newblock \bibinfo{booktitle}{\emph{ChatGPT Prompt Patterns for Improving Code
  Quality, Refactoring, Requirements Elicitation, and Software Design}}.
\newblock \bibinfo{publisher}{Springer Nature Switzerland},
  \bibinfo{address}{Cham}, \bibinfo{pages}{71--108}.
\newblock
\showISBNx{978-3-031-55642-5}
\href{https://doi.org/10.1007/978-3-031-55642-5_4}{doi:\nolinkurl{10.1007/978-3-031-55642-5_4}}


\bibitem[Xiao et~al\mbox{.}(2013)]%
        {xiao_characteristic_2013}
\bibfield{author}{\bibinfo{person}{Xusheng Xiao}, \bibinfo{person}{Sihan Li},
  \bibinfo{person}{Tao Xie}, {and} \bibinfo{person}{Nikolai Tillmann}.}
  \bibinfo{year}{2013}\natexlab{}.
\newblock \showarticletitle{Characteristic studies of loop problems for
  structural test generation via symbolic execution}. In
  \bibinfo{booktitle}{\emph{2013 28th {IEEE}/{ACM} {International} {Conference}
  on {Automated} {Software} {Engineering} ({ASE})}}. \bibinfo{pages}{246--256}.
\newblock
\href{https://doi.org/10.1109/ASE.2013.6693084}{doi:\nolinkurl{10.1109/ASE.2013.6693084}}


\bibitem[Xie et~al\mbox{.}(2023)]%
        {xie_chatunitest_2023}
\bibfield{author}{\bibinfo{person}{Zhuokui Xie}, \bibinfo{person}{Yinghao
  Chen}, \bibinfo{person}{Chen Zhi}, \bibinfo{person}{Shuiguang Deng}, {and}
  \bibinfo{person}{Jianwei Yin}.} \bibinfo{year}{2023}\natexlab{}.
\newblock \bibinfo{title}{ChatUniTest: a ChatGPT-based automated unit test
  generation tool}.
\newblock
\showeprint[arxiv]{2305.04764}~[cs.SE]


\bibitem[Yang et~al\mbox{.}(2024)]%
        {yang_evaluation_2024}
\bibfield{author}{\bibinfo{person}{Lin Yang}, \bibinfo{person}{Chen Yang},
  \bibinfo{person}{Shutao Gao}, \bibinfo{person}{Weijing Wang},
  \bibinfo{person}{Bo Wang}, \bibinfo{person}{Qihao Zhu}, \bibinfo{person}{Xiao
  Chu}, \bibinfo{person}{Jianyi Zhou}, \bibinfo{person}{Guangtai Liang},
  \bibinfo{person}{Qianxiang Wang}, {and} \bibinfo{person}{Junjie Chen}.}
  \bibinfo{year}{2024}\natexlab{}.
\newblock \showarticletitle{On the Evaluation of Large Language Models in Unit
  Test Generation}. In \bibinfo{booktitle}{\emph{Proceedings of the 39th
  IEEE/ACM International Conference on Automated Software Engineering}}
  (Sacramento, CA, USA) \emph{(\bibinfo{series}{ASE '24})}.
  \bibinfo{publisher}{Association for Computing Machinery},
  \bibinfo{address}{New York, NY, USA}, \bibinfo{pages}{1607–1619}.
\newblock
\showISBNx{9798400712487}
\href{https://doi.org/10.1145/3691620.3695529}{doi:\nolinkurl{10.1145/3691620.3695529}}


\bibitem[Yuan et~al\mbox{.}(2024)]%
        {yuan_no_2024}
\bibfield{author}{\bibinfo{person}{Zhiqiang Yuan}, \bibinfo{person}{Mingwei
  Liu}, \bibinfo{person}{Shiji Ding}, \bibinfo{person}{Kaixin Wang},
  \bibinfo{person}{Yixuan Chen}, \bibinfo{person}{Xin Peng}, {and}
  \bibinfo{person}{Yiling Lou}.} \bibinfo{year}{2024}\natexlab{}.
\newblock \showarticletitle{Evaluating and Improving ChatGPT for Unit Test
  Generation}.
\newblock \bibinfo{journal}{\emph{Proc. ACM Softw. Eng.}} \bibinfo{volume}{1},
  \bibinfo{number}{FSE}, Article \bibinfo{articleno}{76} (\bibinfo{date}{July}
  \bibinfo{year}{2024}), \bibinfo{numpages}{24}~pages.
\newblock
\href{https://doi.org/10.1145/3660783}{doi:\nolinkurl{10.1145/3660783}}


\bibitem[Zhang et~al\mbox{.}(2024)]%
        {zhang_assessing_2024}
\bibfield{author}{\bibinfo{person}{Peng Zhang}, \bibinfo{person}{Yang Wang},
  \bibinfo{person}{Xutong Liu}, \bibinfo{person}{Zeyu Lu},
  \bibinfo{person}{Yibiao Yang}, \bibinfo{person}{Yanhui Li},
  \bibinfo{person}{Lin Chen}, \bibinfo{person}{Ziyuan Wang},
  \bibinfo{person}{Chang-Ai Sun}, \bibinfo{person}{Xiao Yu}, {and}
  \bibinfo{person}{Yuming Zhou}.} \bibinfo{year}{2024}\natexlab{}.
\newblock \showarticletitle{Assessing {Effectiveness} of {Test} {Suites}:
  {What} {Do} {We} {Know} and {What} {Should} {We} {Do}?}
\newblock \bibinfo{journal}{\emph{ACM Transactions on Software Engineering and
  Methodology}} \bibinfo{volume}{33}, \bibinfo{number}{4} (\bibinfo{date}{May}
  \bibinfo{year}{2024}), \bibinfo{pages}{1--32}.
\newblock
\showISSN{1049-331X, 1557-7392}
\href{https://doi.org/10.1145/3635713}{doi:\nolinkurl{10.1145/3635713}}


\end{thebibliography}

\end{document}